\documentclass[12pt,a4paper]{article}

\usepackage{graphics}
\usepackage{amsmath}
\usepackage{amssymb}

\title{Mass constraints, production cross sections, and decay rates
in the Two Higgs Doublet Model of type II}
\author{Carlos A. Mar\'{\i}n and B. Hoeneisen}
\date{\small{Universidad San Francisco de Quito \\
5 February 2004}}
\begin{document}
\maketitle

\begin{abstract}
\noindent
We calculate masses, production cross sections, and decay rates
in the Two Higgs Doublet Model of type II.
We also discuss running coupling constants and Grand Unification.
The most interesting production channels are
$gg \rightarrow h^0, H^0, A^0$ on mass shell, and
$q \bar{q}, g g \rightarrow h^0 Z$ and
$q \bar{q'} \rightarrow h^0 W^\pm$ 
in the continuum
(tho there may be peaks at $m_{A^0}$).
The most interesting decays are
$h^0, H^0, A^0 \rightarrow b \bar{b}$-jets and
$\tau^+ \tau^-$, and, if above threshold,
$H^0 \rightarrow 
Z Z$, $W^+ W^-$ and $h^0 h^0$.
The following final states should be compared with the 
Standard Model cross section:
$b \bar{b} Z$, $b \bar{b} W^\pm$, 
$\tau^+ \tau^- Z$, $\tau^+ \tau^- W^\pm$,
$b \bar{b}$, $\tau^+ \tau^-$, $Z Z$,
$W^+ W^-$, 3 and 4 $b$-jets, $2 \tau^+ + 2 \tau^-$,
$b \bar{b} \tau^+ \tau^-$, $Z W^+ W^-$, $3 Z$,
$Z Z W^\pm$ and $3 W^\pm$.
Mass peaks should be searched in the following
channels:
$Z b \bar{b}$, $Z Z$, $Z Z Z$, $b \bar{b}$, $4 b$-jets 
and, just in case, $Z \gamma$.
\end{abstract}

\tableofcontents

\section{Introduction}
Among the extensions of the Standard Model that respect its
principles and symmetries, and are compatible with present data
within a region of parameter space, and are of interest at the
large particle colliders, is the addition of a second doublet of
higgs fields.
In this article we consider the Two Higgs Doublet
Model of type II\cite{1}.
The higgs sector of the Minimal Supersymmetric Standard Model
(MSSM) is of this type (tho the model of type II does not require 
Supersymmetry). 
The physical spectrum of the model
contains five higgs bosons: one pseudoscalar $A^{o}$ (CP-odd
scalar), two neutral scalars $H^{o}$ and $h^{o}$ (CP-even scalars), and
two charged scalars $H^{+}$ and $H^{-}$. The masses of the charged Higgs
bosons $m_H$, and the ratio of the vacuum expectation values of the two
neutral components of the Higgs doublets, $\tan \beta > 0$, are free
parameters of the theory.

In \cite{TwoHiggs} we obtained limits in the $(m_H, \tan\beta)$ plane
using measured decay rates, mixing and CP violation of mesons.
In this article we present 
graphically the corresponding limits on $m_{H^0}$, $m_{h^0}$ 
and $m_{A^0}$. Then we calculate production cross sections,
decay rates and branching fractions of the
higgs particles. Next, we obtain the running coupling constants
and discuss Grand Unification.
Finally, in the Conclusions, we list interesting
discovery channels.

\section{Masses}
The masses of the neutral higgs particles as a function of the
masses of the charged higgs $m_H$, $\tan\beta$ 
and the masses of $Z$ and $W$, calculated at tree level, are:
\begin{equation}
m_{A^0}^2 = m_H^2 - m_W^2,
\label{A}
\end{equation}

\begin{eqnarray}
2 m_{H^0}^2 & = &  m_H^2 - m_W^2 + m_Z^2
\nonumber \\
& & + \left[ \left( m_H^2 - m_W^2 + m_Z^2 \right)^2
- 4 m_Z^2 \left( m_H^2 - m_W^2 \right)
\left( \frac{\tan^2\beta - 1}{\tan^2\beta + 1} 
\right)^2 \right]^{\frac{1}{2}}
\label{H0}
\end{eqnarray}

\begin{eqnarray}
2 m_{h^0}^2 & = & m_H^2 - m_W^2 + m_Z^2
\nonumber \\
& & - \left[ \left( m_H^2 - m_W^2 + m_Z^2 \right)^2
- 4 m_Z^2 \left( m_H^2 - m_W^2 \right)
\left( \frac{\tan^2\beta - 1}{\tan^2\beta + 1} 
\right)^2 \right]^{\frac{1}{2}}
\label{h0}
\end{eqnarray}
We have re derived these equations in agreement with
the literature.\cite{1}

\begin{figure}
\begin{center}
%\vspace*{-4.5cm}
\scalebox{0.8}
{\includegraphics{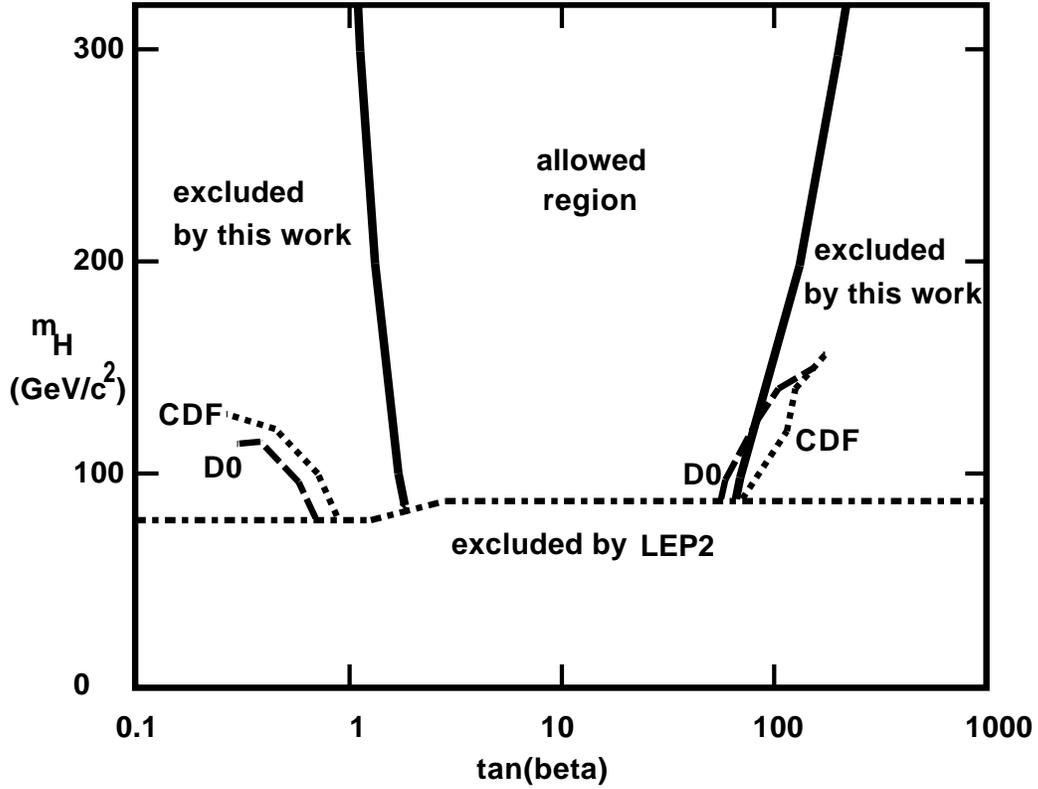}}
%\vspace*{0.7cm}
\caption{Lower and upper limits on $\tan\beta$ as a
function of the mass of the charged higgs $m_H$
from meson decay, mixing and CP violation
(continuous curve) compared
to limits obtained by CDF\cite{CDF}, D0\cite{D0} and LEP2\cite{LEP2}, all
at $95\%$ confidence. Taken from \cite{TwoHiggs}.}
\label{limits_fig}
\end{center}
\end{figure}

\begin{figure}
\begin{center}
%\vspace*{-4.5cm}
\scalebox{0.5}
{\includegraphics{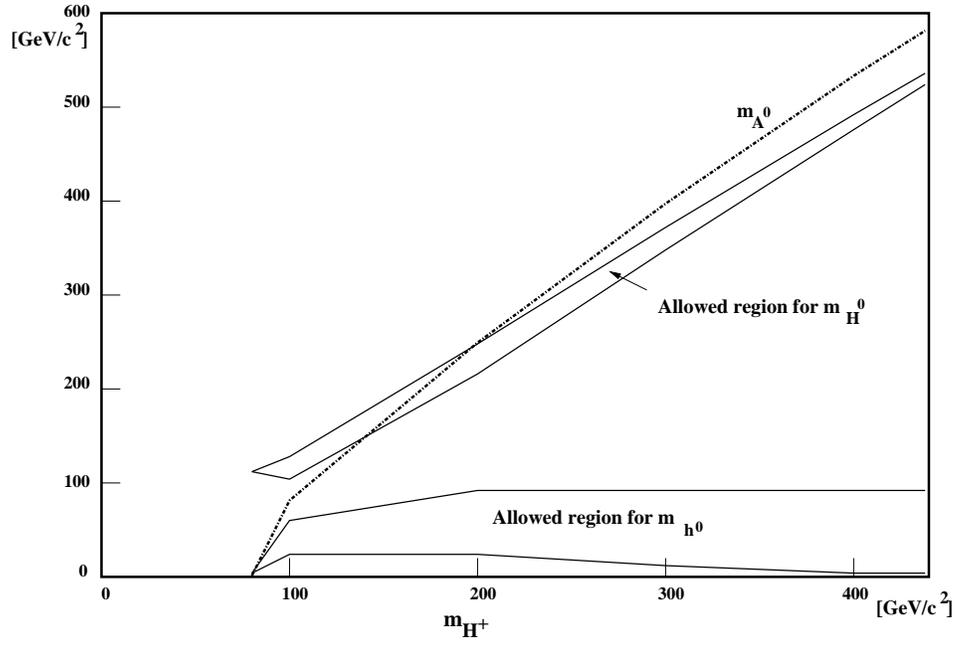}}
%\vspace*{0.7cm}
\caption{Allowed regions of the masses of the neutral
higgs $h^0$, $H^0$ and $A^0$ as a function of the mass
$m_H$ of the charged higgs $H^\pm$.
From Figure \ref{limits_fig} and the tree level Equations \ref{A},
\ref{H0} and \ref{h0}. Radiative corrections raise the
allowed region of $h^0$.\cite{PDG}}
\label{fig3m}
\end{center}
\end{figure}

In reference \cite{TwoHiggs} we obtained the limits
in the $(\tan\beta, m_H)$ plane
shown in Figure \ref{limits_fig}. From that figure and Equations
\ref{A}, \ref{H0} and \ref{h0}
we obtain the limits on the masses of the neutral higgs particles
shown in Figure \ref{fig3m}.

Radiative corrections can be very large. In the MSSM the largest
contributions arise from the incomplete cancellation between top and stop
loops. The corresponding plot similar to 
Figure \ref{fig3m} with radiative corrections
can be found in \cite{PDG}.


\section{Feynman rules}
The Lagrangian for the $V H H$ interaction is:\cite{1}
\begin{eqnarray}
\mathcal{L}_{VHH} & = &
\frac{-ig}{2} W_\mu^+ 
\cdot H^- \overleftrightarrow{\partial}^\mu
\left[ H^0 \sin(\alpha - \beta) + h^0 \cos(\alpha - \beta) + i A^0 \right] 
+ \mathrm{h.c.}
\nonumber \\ & &
- \frac{ig}{2 \cos\theta_W} Z_\mu
\{ i A^0 \overleftrightarrow{\partial}^\mu
\left[ H^0 \sin(\alpha - \beta) + h^0 \cos(\alpha - \beta) \right]
\nonumber \\ & &
- \left( 2 \sin^2\theta_W - 1 \right) 
\cdot H^- \overleftrightarrow{\partial}^\mu H^+
\}
\label{VHH}
\end{eqnarray}
where
\begin{equation}
A \overleftrightarrow{\partial}^\mu B =
A(\partial^\mu B) - (\partial^\mu A) B.
\label{AB}
\end{equation}
The Lagrangian for the $VVH$ interaction is:
\begin{eqnarray}
\mathcal{L}_{VVH} & = &
\left( g m_W W_\mu^+ W^{-\mu} +
\frac{g m_Z}{2 \cos\theta_W} Z_\mu Z^\mu \right)
\nonumber \\ & & 
\times
\left[ H^0 \cos(\beta - \alpha) + h^0 \sin(\beta - \alpha) \right].
\label{VVH}
\end{eqnarray}
There are no vertices $Z H^0 H^0$, 
$Z h^0 h^0$, $Z A^0 A^0$,
$Z W^+ H^-$ or $Z H^0 h^0$. 
The interactions of neutral higgs bosons with up and down quarks
are given by:
\begin{eqnarray}
\mathcal{L}_{AHhff'} & = &
\frac{-g m_f}{2 m_W \sin\beta}
\left[ \bar{u}_f v_{\bar{f}} ( H^0 \sin\alpha + h^0 \cos\alpha) 
- i \cos\beta \cdot \bar{u}_f \gamma^5 v_{\bar{f}} A^0 \right]
\nonumber \\ & &
- \frac{g m_{f'}}{2 m_W \cos\beta}
\nonumber \\ & &
\times
\left[ \bar{u}_{f'} v_{\bar{f'}} (H^0 \cos\alpha - h^0 \sin\alpha )
- i \sin\beta \cdot \bar{u}_{f'} \gamma^5 v_{\bar{f'}} A^0 \right]
\label{hff}
\end{eqnarray}
where $f = u, c, t, \nu_e, \nu_\mu, \nu_\tau$ and
$f' = d, s, b, e^-, \mu^-, \tau^-$.
The Lagrangian corresponding to the $H^{\pm} f \bar{f}'$ vertex is:
\begin{eqnarray}
\mathcal{L}_{Hff'} & = &
\frac{g}{2 \sqrt{2} m_{W}} \{ H^{+} V_{f f'} \bar{u}_{f} \left(
A + B \gamma^{5} \right) v_{\bar{f}'} 
\nonumber \\ & &
+ H^- V_{f f'}^* \bar{u}_{f'} (A - B \gamma^5 ) v_{\bar{f}} \}
\label{L}
\end{eqnarray}
where
$A \equiv \left( m_{f'} \tan \beta + m_{f} \cot \beta \right)$
and $B \equiv \left( m_{f'} \tan \beta - m_{f} \cot \beta \right)$.
$V_{f f'}$ is  an element of the CKM matrix.
The Lagrangian corresponding to three higgs bosons is:
\begin{eqnarray}
\mathcal{L}_{3h} & = &
-g H^0 \{ H^+ H^- \left[ m_W \cos(\beta - \alpha)
- \frac{m_Z}{2 \cos\theta_W} \cos(2 \beta) \cos(\beta + \alpha) \right]
\nonumber \\ & &
+ \frac{m_Z}{4 \cos\theta_W} H^0 H^0 \cos(2 \alpha) \cos(\beta + \alpha)
\nonumber \\ & &
+ \frac{m_Z}{4 \cos\theta_W} h^0 h^0 
\left[ 2 \sin(2 \alpha) \sin(\beta + \alpha) - \cos(\beta 
+ \alpha) \cos(2\alpha) \right]
\nonumber \\ & &
- \frac{m_Z}{4 \cos\theta_W} A^0 A^0 \cos(2 \beta) \cos(\beta + \alpha)
\}
\nonumber \\ & &
-g h^0 \{
H^+ H^- \left[ m_W \sin(\beta - \alpha) 
+ \frac{m_Z}{2 \cos\theta_W} \cos(2 \beta) \sin(\beta + \alpha) \right]
\nonumber \\ & &
+ \frac{m_Z}{4 \cos\theta_W} h^0 h^0 \cos(2 \alpha) \sin(\beta + \alpha)
\nonumber \\ & &
- \frac{m_Z}{4 \cos\theta_W} H^0 H^0 \left[ 2 \sin(2\alpha) \cos(\beta+\alpha)
+ \sin(\beta+\alpha) \cos(2 \alpha) \right]
\nonumber \\ & &
+ \frac{m_Z}{4 \cos\theta_W} A^0 A^0 \cos(2 \beta) \sin(\beta + \alpha)
\}.
\label{3h}
\end{eqnarray}
Vertexes with four partons including two Higgs bosons are
\begin{eqnarray}
\mathcal{L}_{4} & = & e^2 A_\mu A^\mu H^+ H^- +
\frac{e g \cos \left( 2 \theta_W \right)}{\cos \theta_W} 
A_\mu Z^\mu H^+ H^- \nonumber \\
& & - \frac{e g}{2} \sin \left( \beta - \alpha \right) A_\mu W^{\pm \mu} 
H^0 H^\mp + \frac{e g}{2} \cos \left( \beta - \alpha \right)
A_\mu W^{\pm \mu} h^0 H^\mp \nonumber \\
& & \pm \frac{i g e}{2} A_\mu W^{\pm \mu} A^0 H^\mp,
\label{L4}
\end{eqnarray}
The $H^+H^- \gamma$ vertex is
\begin{equation}
\mathcal{L}_{H^+H^- \gamma} = - i g \sin \theta_W A_\mu H^- 
\overleftrightarrow{\partial^\mu} H^+
\label{HHgamma}
\end{equation}
The higgs propagators are: $i / \left( k^{2} - m^{2} + i
\varepsilon \right)$.

\begin{figure}
\begin{center}
%\vspace*{-4.5cm}
%\scalebox{0.5}
{\includegraphics{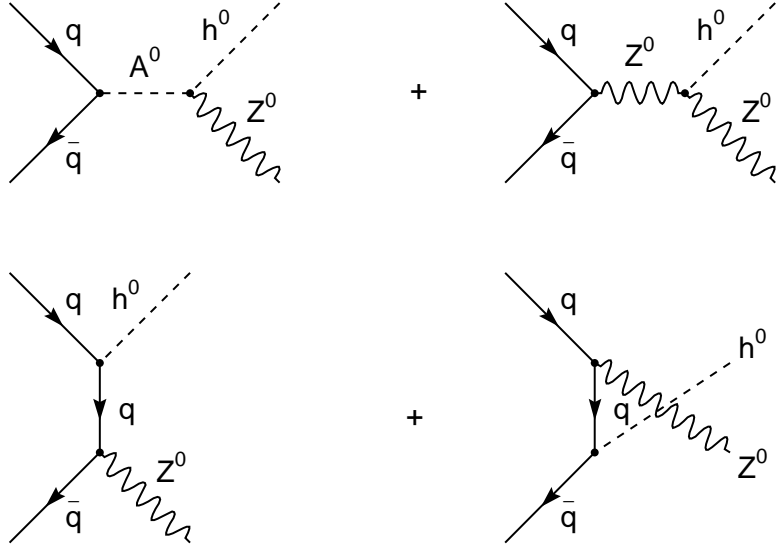}}
%\vspace*{0.7cm}
\caption{Feynman diagrams corresponding to the
production of $h^0$ in the channel
$q \bar{q} \rightarrow h^0 Z^0$.}
\label{nhiggs2}
\end{center}
\end{figure}

Feynman diagrams corresponding to the production of $Z h^0$ are shown
in Figures \ref{nhiggs2}, \ref{nhiggs3} and \ref{nhiggs4}.
Note that the invariant mass of $Z h^0$ can have a resonance at
$m_{A^0}$ which is an interesting experimental signature.
Feynman diagrams corresponding to the production of $W^\pm h^0$ or
$W^\pm H^0$ are shown in Figure \ref{wh_fig}.

\begin{figure}
\begin{center}
%\vspace*{-4.5cm}
%\scalebox{0.5}
{\includegraphics{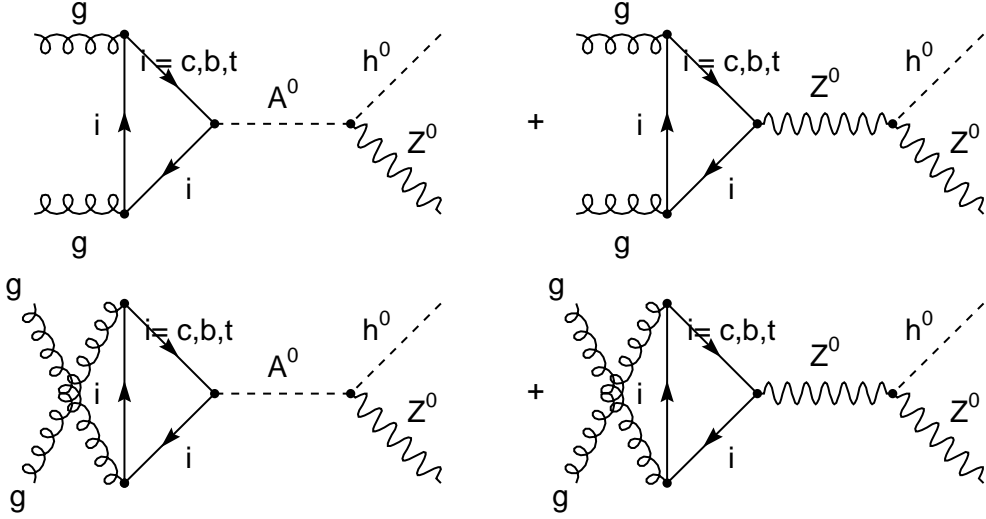}}
%\vspace*{0.7cm}
\caption{Feynman diagrams corresponding to the 
production of $h^0$ in the channel
$g g \rightarrow h^0 Z^0$. Continued in
Figure \ref{nhiggs4}.}
\label{nhiggs3}
\end{center}
\end{figure}
\begin{figure}
\begin{center}
%\vspace*{-4.5cm}
%\scalebox{0.5}
{\includegraphics{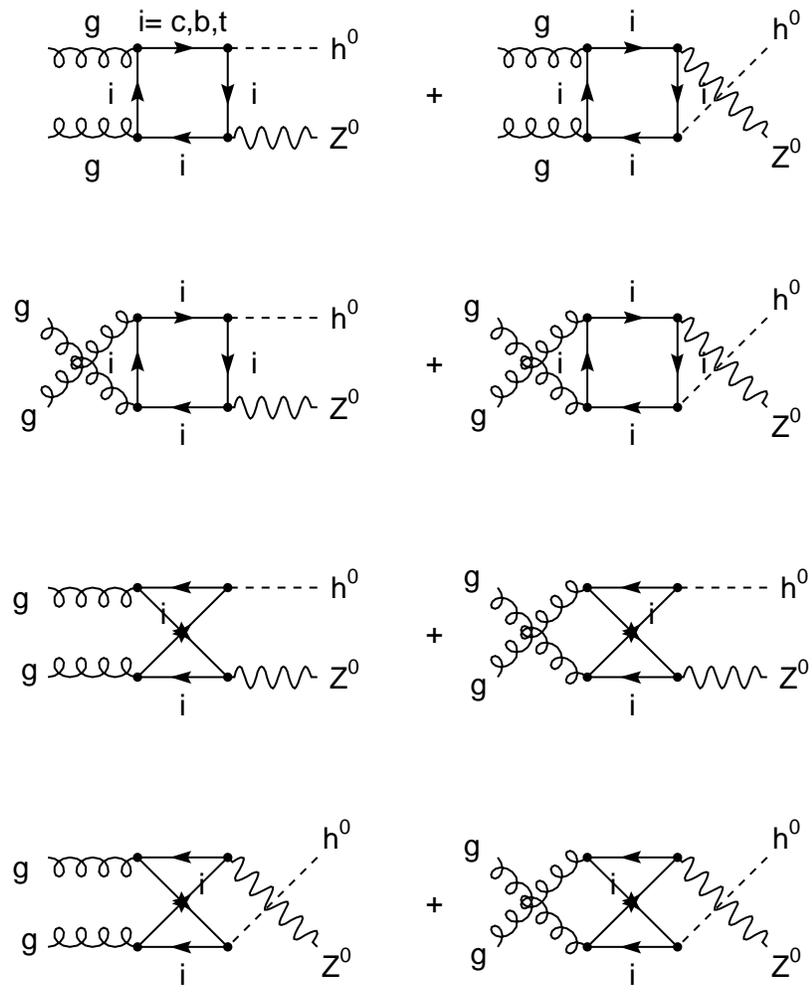}}
%\vspace*{0.7cm}
\caption{Continued from Figure \ref{nhiggs3}.}
\label{nhiggs4}
\end{center}
\end{figure}
\begin{figure}
\begin{center}
%\vspace*{-4.5cm}
%\scalebox{0.5}
{\includegraphics{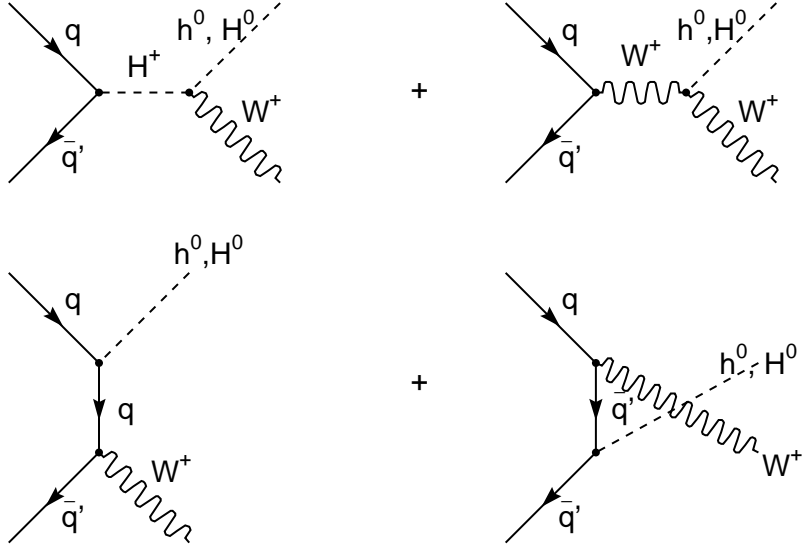}}
%\vspace*{0.7cm}
\caption{Feynman diagrams corresponding to the 
production of $h^0 W^\pm$ and $H^0 W^\pm$.}
\label{wh_fig}
\end{center}
\end{figure}

\section{Decay rates of $h^0$}
\begin{figure}
\begin{center}
%\vspace*{-4.5cm}
%\scalebox{0.5}
{\includegraphics{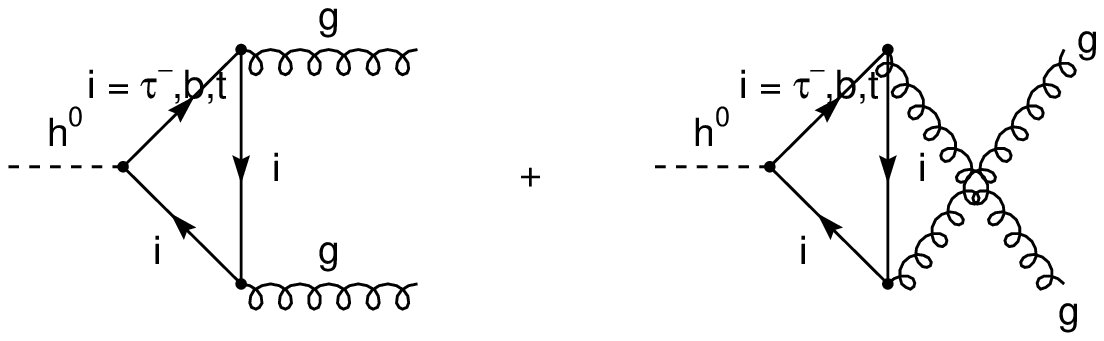}}
%\vspace*{0.7cm}
\caption{Feynman diagrams corresponding to the decay
$h^0 \rightarrow g g$.}
\label{nhiggs}
\end{center}
\end{figure}

Calculating the Feynman diagrams of Figure \ref{nhiggs} we obtain the
decay rate corresponding to $h^0 \rightarrow g g$:
\begin{eqnarray}
\Gamma(h^0 \rightarrow g g) & = &
\frac{\sqrt{2} G_F \alpha_s^2 m_{h^0}^3}{64 \pi^3}
\frac{\cos^2{\alpha}}{\sin^2\beta}
| \tau_b \tan\beta \tan\alpha \left[ \left( \tau_b - 1 \right)
f(\tau_b) + 2 \right] \nonumber \\ & &
+ \tau_\tau \tan\beta \tan\alpha \left[ \left( \tau_\tau - 1 \right)
f(\tau_\tau) + 2 \right] \nonumber \\ & &
- \tau_t \left[ \left( \tau_t - 1 \right) 
f(\tau_t) + 2 \right] | ^2
\label{2g}
\end{eqnarray}
where
\begin{equation}
\tau_i = \frac{4 m_i^2}{m_{h^0}^2},
\label{tau}
\end{equation}
(note that $\tau_i$ will change from Section to Section),
\begin{equation}
f(\tau_i) = \left\{ \begin{array}{ll}
-2 \left[ \arcsin \left( \tau_i^{-1/2} \right) \right]^2 &
\mbox{if $\tau_i > 1$} \\
\frac{1}{2} \left[ \ln \left( \frac{1 + \left( 1- \tau_i \right)^{1/2}}
{1 - \left( 1 - \tau_i \right)^{1/2}} \right) - i \pi \right]^2 &
\mbox{if $\tau_i \le 1$,}
\end{array} \right. 
\label{f}
\end{equation}
\begin{equation}
\tan\alpha = \left\{ \frac{1 + F}{1 - F} \right\}^\frac{1}{2},
\label{tan_alpha}
\end{equation}
\begin{equation}
F = \frac{1 - \tan^2\beta}{\left( 1 + \tan^2\beta \right) G}
\left[ 1 - \frac{m_Z^2}{m_H^2} - \frac{m_W^2}{m_H^2} \right],
\label{F}
\end{equation}
\begin{equation}
G = \left[ \left( 1 + \frac{m_Z^2}{m_H^2} - \frac{m_W^2}{m_H^2} \right)^2
- 4 \frac{m_Z^2}{m_H^2} 
\left( 1 - \frac{m_W^2}{m_H^2} \right)
\left( \frac{\tan^2\beta - 1}{\tan^2\beta + 1} \right)^2
\right]^\frac{1}{2}.
\label{G}
\end{equation}

\begin{figure}
\begin{center}
%\vspace*{-4.5cm}
%\scalebox{0.5}
{\includegraphics{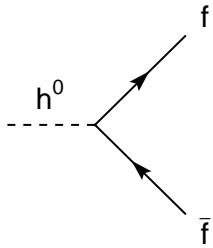}}
%\vspace*{0.7cm}
\caption{Feynman diagram corresponding to the decays
$h^0 \rightarrow b \bar{b}, c \bar{c}, \tau \bar{\tau}$.}
\label{nhiggs5}
\end{center}
\end{figure}

Calculating the Feynman diagrams of Figure \ref{nhiggs5} we obtain:
\begin{equation}
\Gamma\left( h^0 \rightarrow c \bar{c} \right) =
\frac{3 G_F m_c^2 m_{h^0} \cos^2\alpha}{\sqrt{2} \cdot 
4 \pi \sin^2\beta}
\left( 1 - \frac{4 m_c^2}{m_{h^0}^2} \right)^\frac{3}{2},
\label{ccbar}
\end{equation}
\begin{equation}
\Gamma\left( h^0 \rightarrow b \bar{b} \right) =
\frac{3 G_F m_b^2 m_{h^0} \sin^2\alpha}{\sqrt{2} \cdot 
4 \pi \cos^2\beta}
\left( 1 - \frac{4 m_b^2}{m_{h^0}^2} \right)^\frac{3}{2},
\label{bbbar}
\end{equation}
and
\begin{equation}
\Gamma\left( h^0 \rightarrow \tau^- \tau^+ \right) =
\frac{G_F m_\tau^2 m_{h^0} \sin^2\alpha}{\sqrt{2} \cdot 
4 \pi \cos^2\beta}
\left( 1 - \frac{4 m_\tau^2}{m_{h^0}^2} \right)^\frac{3}{2}.
\label{tautaubar}
\end{equation}


\section{Branching fractions of $h^0$}
From the preceding decay rates we obtain the following
branching fractions for the case
$m_b, m_c, m_\tau \ll m_{h^0} < 120$GeV/c$^2$:
\begin{equation}
B \left( h^0 \rightarrow b \bar{b} \right) =
\frac{3 m_b^2 \sin^2\alpha}{3 m_b^2 \sin^2\alpha + 
3 m_c^2 \cos^2\alpha \cot^2\beta 
+ m_\tau^2 \sin^2\alpha + J}
\label{Bbbar}
\end{equation}
and
\begin{equation}
B \left( h^0 \rightarrow \tau^+ \tau^- \right) =
\frac{m_\tau^2 \sin^2\alpha}{3 m_b^2 \sin^2\alpha + 
3 m_c^2 \cos^2\alpha \cot^2\beta 
+ m_\tau^2 \sin^2\alpha + J}
\label{Btau}
\end{equation}
where
\begin{eqnarray}
J & = & \frac{\alpha_s^2 m_{h^0}^2}{8\pi^2} \frac{\cos^2\alpha}{\tan^2\beta}
\vert 
\tan\beta \tan\alpha \sum_{j=b,\tau}{\left[ \tau_j \left( -
\frac{1}{2} \left\{ \ln \left( \frac{\tau_j}{4} \right)
+ i\pi \right\}^2 + 2 \right) \right]} \nonumber \\ & &
- \tau_t \left\{ \left( \tau_t - 1 \right) 
f \left( \tau_t \right) + 2 \right\} \vert^2.
\label{J}
\end{eqnarray}
For $90 < m_H < 1000$GeV/c$^2$, $B \left( h^0 \rightarrow b \bar{b} \right)$
varies from 0.856 to 0.944. Neglecting $B\left( h^0 \rightarrow g g \right)$
and the contribution of $c \bar{c}$ we obtain
\begin{equation}
B \left( h^0 \rightarrow b \bar{b} \right) =
\frac{3m_b^2}{3m_b^2 + m_\tau^2} = 0.944
\label{Bbbar_short}
\end{equation}
and
\begin{equation}
B \left( h^0 \rightarrow \tau^+ \tau^- \right) =
\frac{m_\tau^2}{3 m_b^2 + m_\tau^2} = 0.056.
\label{Btau_short}
\end{equation}

\section{Decay rates of $H^{\pm}$}
\begin{figure}
\begin{center}
%\vspace*{-4.5cm}
%\scalebox{0.5}
{\includegraphics{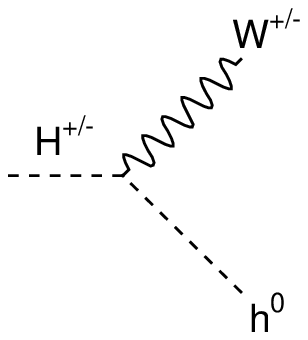}}
%\vspace*{0.7cm}
\caption{Feynman diagram corresponding to the decay
$H^\pm \rightarrow W^\pm h^0$.}
\label{H_Wh_fig}
\end{center}
\end{figure}

\begin{figure}
\begin{center}
%\vspace*{-4.5cm}
%\scalebox{0.5}
{\includegraphics{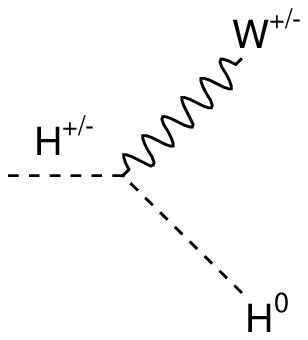}}
%\vspace*{0.7cm}
\caption{Feynman diagram corresponding to the decay
$H^\pm \rightarrow W^\pm H^0$.}
\label{H__WH_fig}
\end{center}
\end{figure}

\begin{figure}
\begin{center}
%\vspace*{-4.5cm}
%\scalebox{0.5}
{\includegraphics{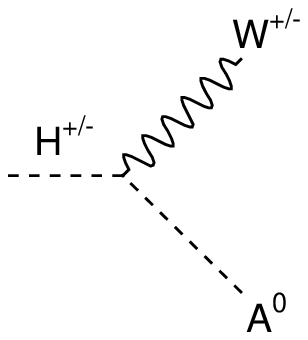}}
%\vspace*{0.7cm}
\caption{Feynman diagram corresponding to the decay
$H^\pm \rightarrow W^\pm A^0$.}
\label{Hpm_Wpm_A_fig}
\end{center}
\end{figure}

The tree level Feynman diagram of Figure \ref{H_Wh_fig} gives 
the following decay rate:
\begin{eqnarray}
\Gamma \left( H^{\pm} \rightarrow W^{\pm} h^0 \right) & = &
\frac{\sqrt{2} G_F \cos^2{\alpha}}
{16 \pi m_H^3 \left( 1 + \tan^2{\beta} \right)}
\left[ 1 + \tan{\beta} \tan{\alpha} \right]^2 
\nonumber \\ & &
\times\Lambda^{3/2} \left(m_H^2, m_W^2, m_{h^0}^2 \right)
\label{H_Wh}
\end{eqnarray}
where
\begin{equation}
\Lambda(a, b, c) = a^2 + b^2 + c^2 - 2ab - 2bc - 2ca.
\label{Lambda}
\end{equation}
Similarly from the Feynman diagrams of Figures \ref{H__WH_fig} 
and \ref{Hpm_Wpm_A_fig} we obtain
\begin{eqnarray}
\Gamma \left( H^{\pm} \rightarrow W^{\pm} H^0 \right) & = &
\frac{\sqrt{2} G_F \left( \tan\beta - \tan\alpha \right)^2 }
{16 \pi m_H^3 \left( 1 + \tan^2{\beta} \right)
\left( 1 + \tan^2{\alpha} \right)}
\nonumber \\ & &
\times\Lambda^{3/2} \left(m_H^2, m_W^2, m_{H^0}^2 \right),
\label{H__WH}
\end{eqnarray}
\begin{equation}
\Gamma \left( H^\pm \rightarrow W^\pm A^0 \right) =
\frac{\sqrt{2} G_F}{16 \pi m_H^3}
\Lambda^{3/2} \left( m_{A^0}^2, m_W^2, m_H^2 \right).
\label{H_WA}
\end{equation}


\section{Decays of $H^0$.}
\begin{figure}
\begin{center}
%\vspace*{-4.5cm}
%\scalebox{0.5}
{\includegraphics{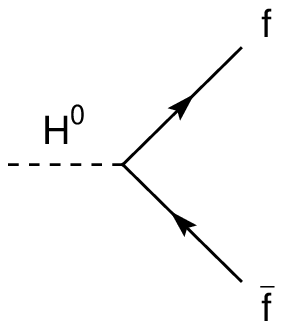}}
%\vspace*{0.7cm}
\caption{Feynman diagram corresponding to the decay
$H^0 \rightarrow f \bar{f}$.}
\label{H0_ff_fig}
\end{center}
\end{figure}

The tree level Feynman diagrams of Figure \ref{H0_ff_fig}
give the following decay rates:
\begin{equation}
\Gamma(H^0 \rightarrow f \bar{f}) =
\frac{\sqrt{2} G_F m_f^2 m_{H^0}}{8 \pi}
\left( 1 - \frac{4 m_f^2}{m_{H^0}^2} \right)^{3/2} N_f B_f^2
\label{H0_ff}
\end{equation}
where $N_f = 3$ for quarks, $N_f = 1$ for leptons,
$B_f^2 = \sin^2\alpha ( 1 + \cot^2\beta )$ for $f = u, c, t$, and
$B_f^2 = \cos^2\alpha ( 1 + \tan^2\beta )$ for $f = d, s, b, 
e^-, \mu^-, \tau^-$. 

\begin{figure}
\begin{center}
%\vspace*{-4.5cm}
%\scalebox{0.5}
{\includegraphics{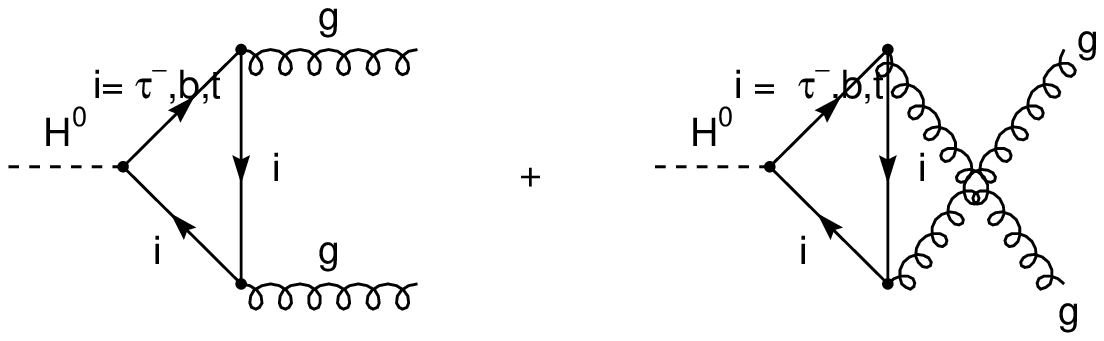}}
%\vspace*{0.7cm}
\caption{Feynman diagrams corresponding to the decay
$H^0 \rightarrow g g$.}
\label{H0_gg_fig}
\end{center}
\end{figure}

\begin{figure}
\begin{center}
%\vspace*{-4.5cm}
%\scalebox{0.5}
{\includegraphics{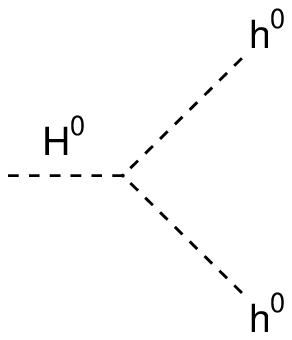}}
%\vspace*{0.7cm}
\caption{Feynman diagram corresponding to the decay
$H^0 \rightarrow h^0 h^0$.}
\label{H_hh_fig}
\end{center}
\end{figure}

\begin{figure}
\begin{center}
%\vspace*{-4.5cm}
%\scalebox{0.5}
{\includegraphics{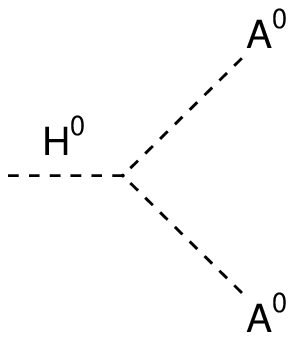}}
%\vspace*{0.7cm}
\caption{Feynman diagram corresponding to the decay
$H^0 \rightarrow A^0 A^0$.}
\label{H_AA_fig}
\end{center}
\end{figure}

\begin{figure}
\begin{center}
%\vspace*{-4.5cm}
%\scalebox{0.5}
{\includegraphics{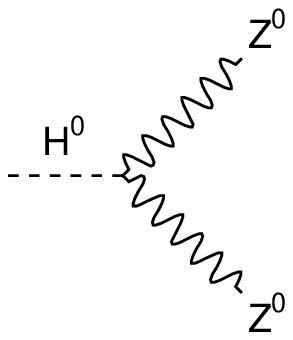}}
%\vspace*{0.7cm}
\caption{Feynman diagram corresponding to the decay
$H^0 \rightarrow Z Z$.}
\label{H0_ZZ_fig}
\end{center}
\end{figure}

From the Feynman diagrams of Figure \ref{H0_gg_fig} we obtain
\begin{eqnarray}
\Gamma(H^0 \rightarrow g g) & = &
\frac{\sqrt{2} G_F \alpha_s^2 m_{H^0}^3}{64 \pi^3}
\sin^2\alpha ( 1+ \cot^2\beta ) \nonumber \\ & &
| \frac{\tan\beta}{\tan\alpha} \sum_{i=b,\tau}{ \tau_i
\left[ ( \tau_i - 1 ) f(\tau_i) + 2 \right]} \nonumber \\ & &
+ { \tau_t \left[ ( \tau_t - 1 ) f(\tau_t) + 2 \right]} |^2 
\label{H_gg}
\end{eqnarray}
where
\begin{equation}
\tau_i = \frac{4 m_i^2}{m_{H^0}^2}.
\label{tau_H0}
\end{equation}
From the Feynman diagram of Figure \ref{H_hh_fig} we obtain
\begin{eqnarray}
\Gamma \left( H^0 \rightarrow h^0 h^0 \right) & = &
\frac{\sqrt{2} G_F m_Z^4}{32 \pi m_{H^0}}
\left( 1 - \frac{4 m_{h^0}^2}{m_{H^0}^2} \right)^{1/2}
\frac{\left( 1 - \tan^2\alpha \right)^2}
{\left( 1 + \tan^2\alpha \right)^3 \left( 1 + \tan^2\beta \right)}
\nonumber \\ & &
\left[ \frac{4 \tan\alpha}{1 - \tan^2\alpha}
\left( \tan\alpha + \tan\beta \right)
- \left( 1 - \tan\alpha \tan\beta \right) \right]^2.
\label{H_hh}
\end{eqnarray}
From the Feynman diagrams of Figures \ref{H_AA_fig} 
and \ref{H0_ZZ_fig} we obtain
\begin{eqnarray}
\Gamma \left( H^0 \rightarrow A^0 A^0 \right) & = &
\frac{\sqrt{2} G_F m_Z^4}{32 \pi m_{H^0}}
\left( 1 - \frac{4 m_{A^0}^2}{m_{H^0}^2} \right)^{1/2}
\nonumber \\ & &
\frac{\left( \tan^2\beta - 1 \right)^2}{ \left( 1 + \tan^2\alpha \right)
\left( 1 + \tan^2\beta \right)^3}
\left[ 1 - \tan\alpha \tan\beta \right]^2,
\label{H_AA}
\end{eqnarray}
\begin{eqnarray}
\Gamma \left( H^0 \rightarrow Z Z \right) & = &
\frac{\sqrt{2} G_F \left( 1 + \tan\beta \tan\alpha \right)^2 m_{H^0}^3}
{32 \pi \left( 1 + \tan^2\beta \right) \left( 1 + \tan^2\alpha \right)}
\nonumber \\ & &
\times \left( 12 x^2 - 4 x + 1 \right) \left( 1 - 4 x \right)^{1/2}
\label{H_ZZ}
\end{eqnarray}
where
\begin{equation}
x = \frac{m_Z^2}{m_{H^0}^2}.
\label{x}
\end{equation}

\begin{figure}
\begin{center}
%\vspace*{-4.5cm}
%\scalebox{0.5}
{\includegraphics{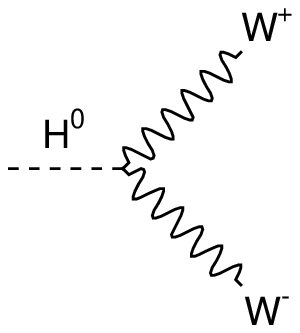}}
%\vspace*{0.7cm}
\caption{Feynman diagram corresponding to the decay
$H^0 \rightarrow W^+ W^-$.}
\label{H0_WW_fig}
\end{center}
\end{figure}

Similarly, from the diagram of Figure \ref{H0_WW_fig} we obtain
\begin{eqnarray}
\Gamma \left( H^0 \rightarrow W^+ W^- \right) & = &
\frac{\sqrt{2} G_F \left( 1 + \tan\beta \tan\alpha \right)^2 m_{H^0}^3}
{16 \pi \left( 1 + \tan^2\beta \right) \left( 1 + \tan^2\alpha \right)}
\nonumber \\ & &
\times \left( 12 y^2 - 4y + 1 \right) \left( 1 - 4 y \right)^{1/2}
\label{H_WW}
\end{eqnarray}
where
\begin{equation}
y = \frac{m_W^2}{m_{H^0}^2}.
\label{y}
\end{equation}

\begin{figure}
\begin{center}
%\vspace*{-4.5cm}
%\scalebox{0.5}
{\includegraphics{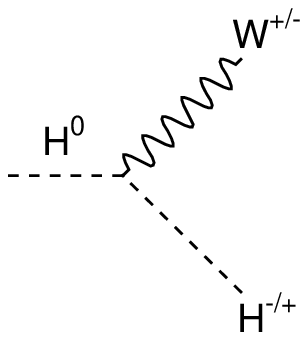}}
%\vspace*{0.7cm}
\caption{Feynman diagram corresponding to the decay
$H^0 \rightarrow W^\pm H^\mp$.}
\label{H_WH_fig}
\end{center}
\end{figure}

From the Feynman diagram of Figure \ref{H_WH_fig} we obtain
\begin{eqnarray}
\Gamma \left( H^0 \rightarrow W^\pm H^\mp \right) & = &
\frac{\sqrt{2} G_F \left( \tan\alpha - \tan\beta \right)^2}
{16 \pi m_{H^0}^3 \left( 1 + \tan^2\alpha \right) \left( 1 + \tan^2\beta \right)}
\nonumber \\ & &
\times \Lambda^{3/2} \left( m_{H^0}^2, m_W^2, m_H^2 \right).
\label{H_WH}
\end{eqnarray}

\begin{figure}
\begin{center}
%\vspace*{-4.5cm}
%\scalebox{0.5}
{\includegraphics{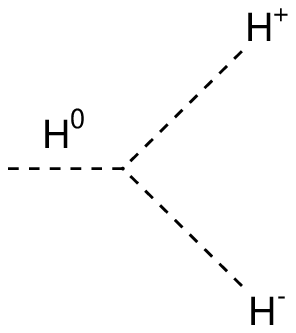}}
%\vspace*{0.7cm}
\caption{Feynman diagram corresponding to the decay
$H^0 \rightarrow H^+ H^-$.}
\label{H_HH_fig}
\end{center}
\end{figure}

From the Feynman diagram of Figure \ref{H_HH_fig} we obtain
\begin{eqnarray}
\lefteqn{
\Gamma \left( H^0 \rightarrow H^+ H^- \right)  = 
\frac{\sqrt{2} G_F m_W^4}{4 \pi m_{H^0} \left( 1 + \tan^2\beta \right)
\left( 1 + \tan^2\alpha \right)}
}
\nonumber \\ & &
\times \left[ \left( 1 + \tan\beta \tan\alpha \right) 
- \frac{ \left(1 - \tan\beta \tan\alpha \right)}{2 \cos^2\theta_W}
\frac{1 - \tan^2\beta}{1 + \tan^2\beta} \right]^2
\nonumber \\ & &
\times \left( 1 - \frac{4 m_H^2}{m_{H^0}^2} \right)^{1/2}.
\label{H_HH}
\end{eqnarray}

\begin{figure}
\begin{center}
%\vspace*{-4.5cm}
%\scalebox{0.5}
{\includegraphics{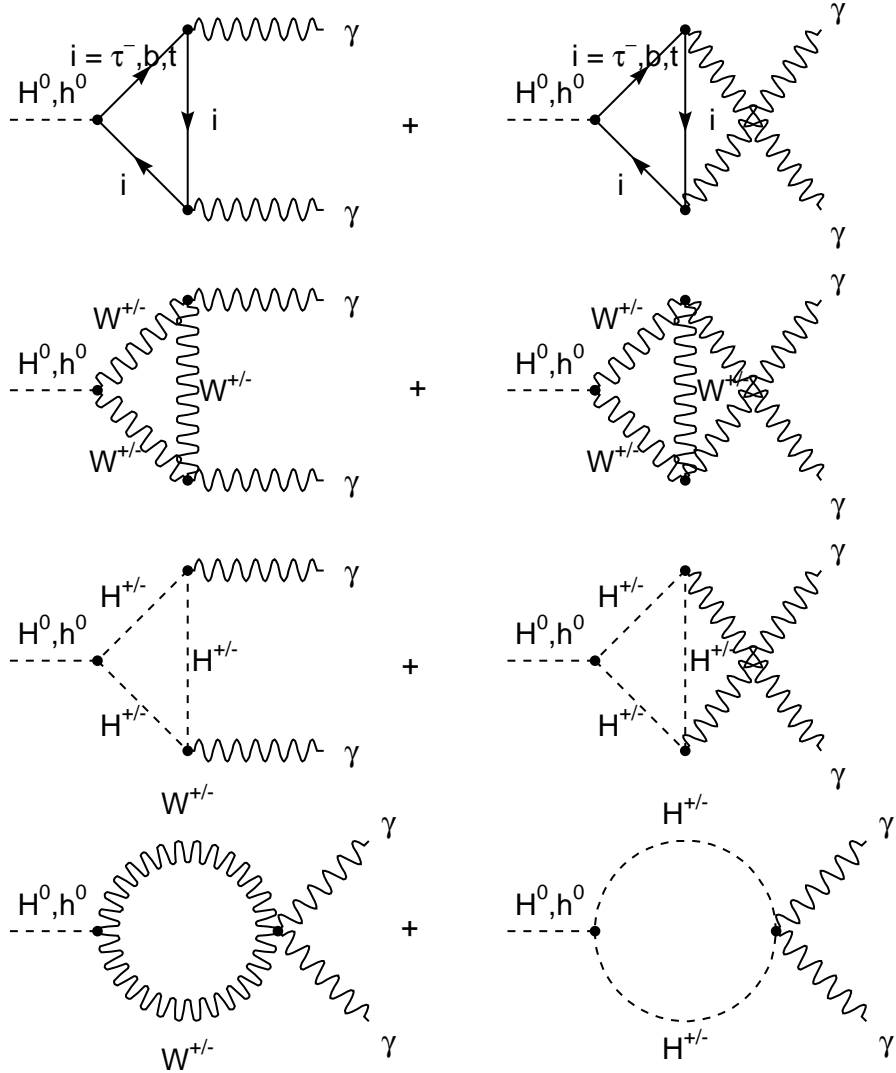}}
%\vspace*{0.7cm}
\caption{Feynman diagrams corresponding
to $h^0, H^0 \rightarrow \gamma \gamma$.}
\label{hhgamma_fig}
\end{center}
\end{figure}

The Feynman diagrams corresponding to 
$h^0, H^0 \rightarrow \gamma \gamma$ are shown in 
Figure \ref{hhgamma_fig}.
\begin{equation}
\Gamma ( H^0 \rightarrow h^0 Z) = 0.
\label{HhZ}
\end{equation}


\section{Decay rates of $A^0$}
\begin{figure}
\begin{center}
%\vspace*{-4.5cm}
%\scalebox{0.5}
{\includegraphics{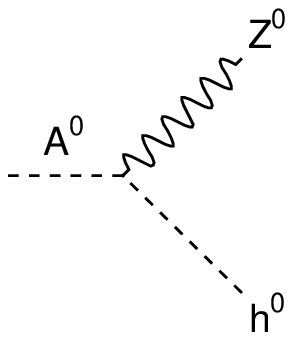}}
%\vspace*{0.7cm}
\caption{Feynman diagram corresponding to the decay
$A^0 \rightarrow Z h^0$.}
\label{A_Zh_fig}
\end{center}
\end{figure}

\begin{figure}
\begin{center}
%\vspace*{-4.5cm}
%\scalebox{0.5}
{\includegraphics{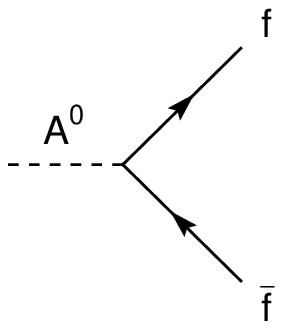}}
%\vspace*{0.7cm}
\caption{Feynman diagram corresponding to the decay
$A^0 \rightarrow f \bar{f}$.}
\label{A_ff_fig}
\end{center}
\end{figure}

From the tree level Feynman diagram of Figure \ref{A_Zh_fig} 
we obtain
\begin{equation}
\Gamma \left( A^0 \rightarrow Z h^0 \right) =
\frac{\sqrt{2} G_F \cos^2{\alpha}}
{16 \pi m_{A^0}^3 \left( 1 + \tan^2{\beta} \right)}
\left[ 1 + \tan{\beta} \tan{\alpha} \right]^2 
\Lambda^{3/2} \left( m_{A^0}^2, m_{h^0}^2, m_Z^2 \right).
\label{A_Zh}
\end{equation}
From the tree level diagram of Figure \ref{A_ff_fig} we obtain
\begin{equation}
\Gamma \left( A^0 \rightarrow f \bar{f} \right) =
\frac{\sqrt{2} G_F}{8 \pi} m_{A^0} m_f^2 A_f^2
\left(1 - \frac{4 m_f^2}{m_{A^0}^2} \right)^{1/2} N_f
\label{A_ff}
\end{equation}
where
$N_f = 3$ for quarks, $N_f = 1$ for leptons,
$A_f = \cot{\beta}$ for $f = u, c, t$, and
$A_f = \tan{\beta}$ for $f = d, s, b, e^-, \mu^-, \tau^-$.

\begin{figure}
\begin{center}
%\vspace*{-4.5cm}
%\scalebox{0.5}
{\includegraphics{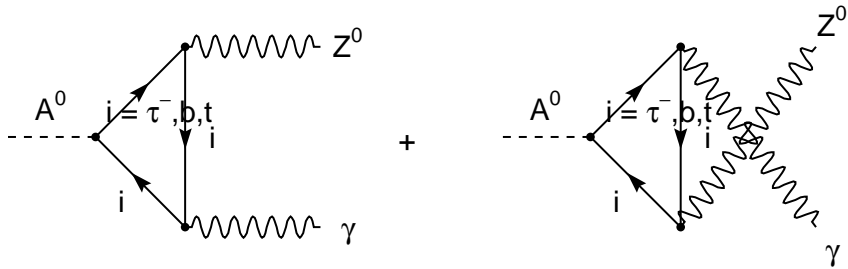}}
%\vspace*{0.7cm}
\caption{Feynman diagrams corresponding to the 
decay $A^0 \rightarrow Z \gamma$.}
\label{A_Zgamma_fig}
\end{center}
\end{figure}

From the Feynman diagrams shown in Figure \ref{A_Zgamma_fig} we
obtain:
\begin{eqnarray}
\Gamma \left( A^0 \rightarrow Z \gamma \right) & = &
\frac{\sqrt{2} G_F \alpha_{em}^2 m_{A^0}^3}{512 \pi^3 \sin^2{\theta_W}
\cos^2{\theta_W}}
\left( 1 - \frac{m_Z^2}{m_{A^0}^2} \right)^3
| \tan{\beta} 
\nonumber \\ & & 
\left[ \left( \frac{1}{2} - \frac{2}{3} \sin^2{\theta_W} \right) 
I \left( \tau_b, \Lambda_b \right)
+ \left( \frac{1}{2} - 2 \sin^2{\theta_W} \right)
I \left( \tau_\tau, \Lambda_\tau \right) \right]
\nonumber \\ & &
+ 2 \cot{\beta} \left( \frac{1}{2} - \frac{4}{3} \sin^2{\theta_W} \right)
I \left( \tau_t, \Lambda_t \right) |^2
\label{A_Zgamma}
\end{eqnarray}
where
\begin{equation}
\tau_i = \frac{4 m_i^2}{m_{A^0}^2}, \qquad
\Lambda_i = \frac{4 m_i^2}{m_Z^2},
\label{Lambda_i}
\end{equation}
and
\begin{equation}
I \left( \tau_i, \Lambda_i \right) =
\frac{\tau_i \Lambda_i}{\Lambda_i - \tau_i}
\left\{ f(\tau_i) - f(\Lambda_i) \right\}.
\label{I}
\end{equation}

\begin{figure}
\begin{center}
%\vspace*{-4.5cm}
%\scalebox{0.5}
{\includegraphics{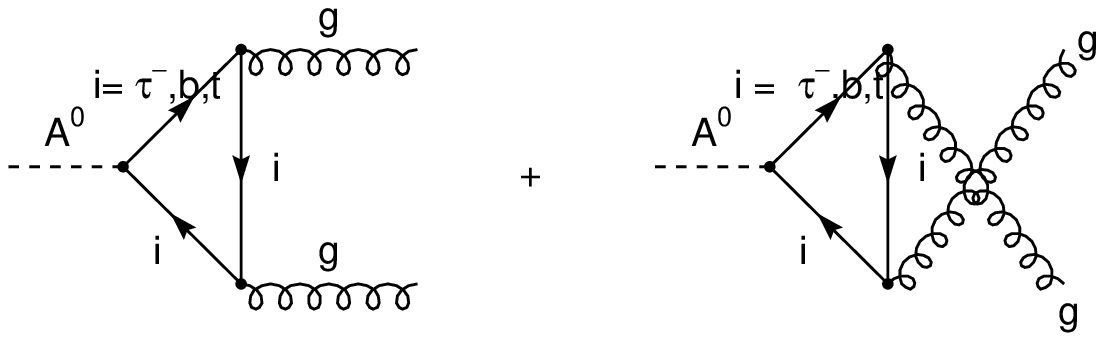}}
%\vspace*{0.7cm}
\caption{Feynman diagrams of $A^0 \rightarrow g g$.}
\label{A_gg}
\end{center}
\end{figure}

From the Feynman diagrams of Figure \ref{A_gg} we obtain the decay rate:
\begin{equation}
\Gamma ( A^0 \rightarrow g g ) =
\frac{\sqrt{2} G_F \alpha_s^2 m_{A^0}^3}{128 \pi^3}
|\tan\beta \sum_{i=b,\tau} \tau_i f(\tau_i) +
\cot\beta \tau_t f(\tau_t) |^2.
\label{A_to_2g}
\end{equation}

\begin{equation}
\Gamma \left( A^0 \rightarrow h^0 h^0 \right) = 
\Gamma \left( A^0 \rightarrow H^0 H^0 \right) = 0.
\label{A_hh}
\end{equation}

\begin{figure}
\begin{center}
%\vspace*{-4.5cm}
%\scalebox{0.5}
{\includegraphics{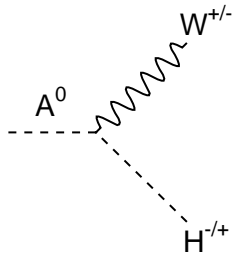}}
%\vspace*{0.7cm}
\caption{Feynman diagram of $A^0 \rightarrow W^\pm H^\mp$.}
\label{A_WH_fig}
\end{center}
\end{figure}

From the Feynman diagram \ref{A_WH_fig} we obtain
\begin{equation}
\Gamma \left( A^0 \rightarrow W^\pm H^\mp \right) =
\frac{\sqrt{2} G_F}{16 \pi m_{A^0}^3}
\Lambda^{3/2} \left(m_{A^0}^2, m_W^2, m_H^2 \right). 
\label{A_WH}
\end{equation}

\begin{figure}
\begin{center}
%\vspace*{-4.5cm}
%\scalebox{0.5}
{\includegraphics{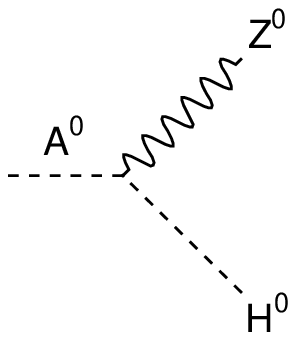}}
%\vspace*{0.7cm}
\caption{Feynman diagram of $A^0 \rightarrow Z H^0$.}
\label{A_ZH_fig}
\end{center}
\end{figure}

From the Feynman diagram \ref{A_ZH_fig} we obtain
\begin{eqnarray}
\Gamma \left( A^0 \rightarrow Z H^0 \right) & = &
\frac{\sqrt{2} G_F \left( \tan\beta - \tan\alpha \right)^2}
{16 \pi m_{A^0}^3 \left( 1 + \tan^2\alpha \right)
\left( 1 + \tan^2\beta \right)}
\nonumber \\ & &
\times \Lambda^{3/2} \left( m_{A^0}^2, m_{H^0}^2, m_Z^2 \right).
\label{A_ZH}
\end{eqnarray}

\begin{figure}
\begin{center}
%\vspace*{-4.5cm}
%\scalebox{0.5}
{\includegraphics{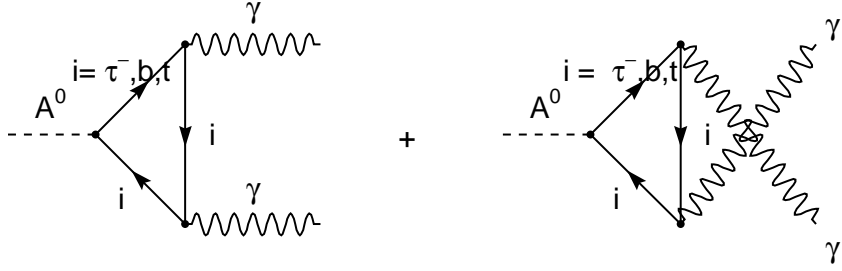}}
%\vspace*{0.7cm}
\caption{Feynman diagrams for $A^0 \rightarrow \gamma \gamma$.}
\label{Agammagamma_fig}
\end{center}
\end{figure}

From the Feynman diagrams of \ref{Agammagamma_fig}
we obtain:
\begin{eqnarray}
\Gamma \left( A^0 \rightarrow \gamma \gamma \right) & = &
\frac{\sqrt{2} G_F \alpha^2_{em} m^3_{A^0}}{256 \pi^3}
\nonumber \\ & & \times
\left| \tan\beta \left[ \frac{1}{3} \tau_b f(\tau_b) +
\tau_\tau f(\tau_\tau) \right]
+ \cot\beta \left[ \frac{4}{3} \tau_t f(\tau_t) \right] \right|^2
\label{Agammagamma}
\end{eqnarray}


\section{Decay $Z \rightarrow h^0 \gamma$}

\begin{figure}
\begin{center}
%\vspace*{-4.5cm}
%\scalebox{0.5}
{\includegraphics{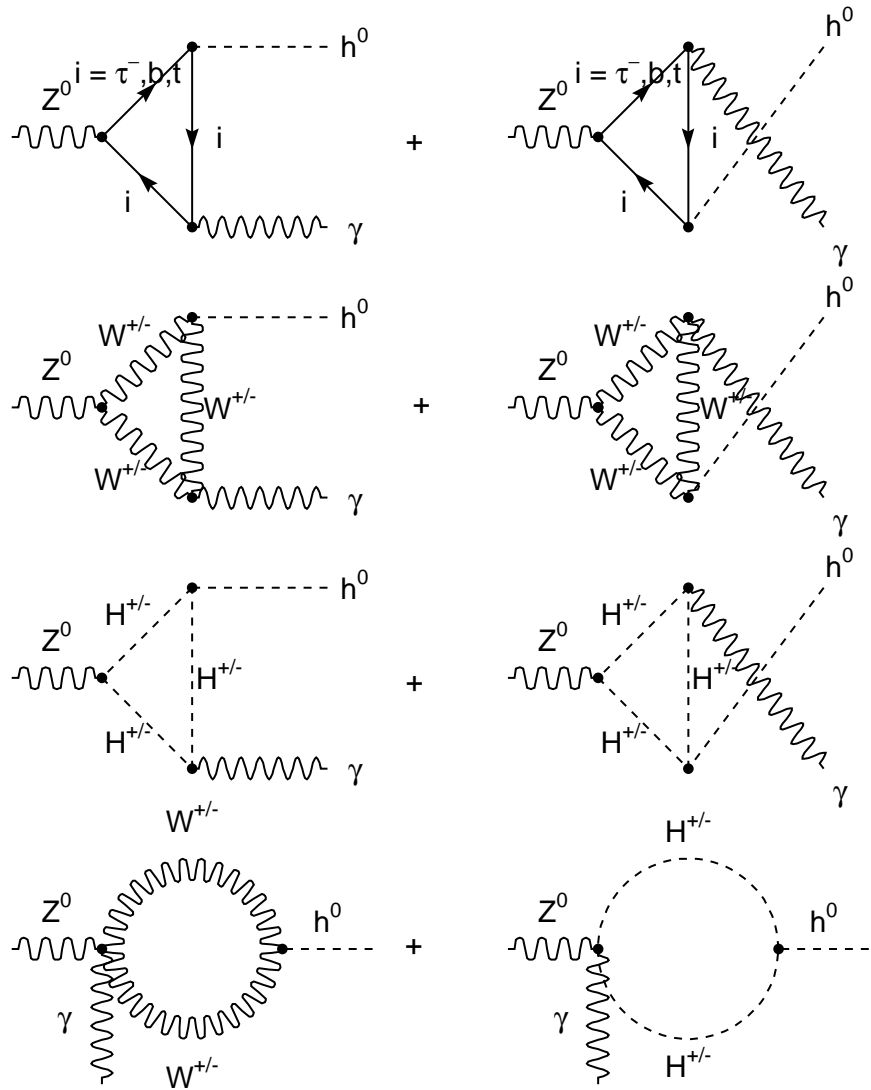}}
%\vspace*{0.7cm}
\caption{Feynman diagrams for $Z \rightarrow h^0 \gamma$.}
\label{Z_hgamma_fig}
\end{center}
\end{figure}

From the Feynman diagrams of Figure \ref{Z_hgamma_fig} we obtain:
\begin{eqnarray}
\Gamma \left( Z \rightarrow h^0 \gamma \right) & = &
\frac{\sqrt{2} G_F \alpha_{em}^2 m_Z^3}{64 \pi^3 \sin^2{\theta_W} \cos^2{\theta_W}}
\left( 1 - \frac{m_{h^0}^2}{m_Z^2} \right)^3 
\cos^2\alpha \cos^2\beta
\nonumber \\ & & 
\mathopen{\vert} \frac{1}{\cos\beta \sin\beta}
[ \tan{\alpha} \tan{\beta} \left( \frac{1}{2}
- \frac{2}{3} \sin^2{\theta_W} \right) F(\tau_b, \Lambda_b)
\nonumber \\ & &
+ \tan{\alpha} \tan{\beta}
\left( \frac{1}{2} - 2 \sin^2{\theta_W} \right) F(\tau_\tau, \Lambda_\tau) 
\nonumber \\ & &
- 2 \left( \frac{1}{2} - \frac{4}{3} \sin^2{\theta_W} \right) 
F(\tau_t, \Lambda_t) ]
\nonumber \\ & &
+ \left[ \tan\beta - \tan\alpha +
\frac{1 - \tan^2\beta}{1 + \tan^2\beta} \frac{\tan\beta + \tan\alpha}
{2 \cos^2\theta_W} \right]
\nonumber \\ & &
\times
\frac{m_W^2}{2 m_H^2} (1 - 2 \sin^2\theta_W) I(\tau_H, \Lambda_H)
\nonumber \\ & &
- \frac{1}{2} \left( \tan \beta - \tan \alpha \right) \cos^2 \theta_W
[ 4 \left( 3 - \tan^2 \theta_W \right) K( \tau_W, \Lambda_W)
\nonumber \\ & &
+ \left\{ \left( 1 + \frac{2}{\tau_W} \right) \tan^2 \theta_W
- \left( 5 + \frac{2}{\tau_W} \right) \right\} I(\tau_W, \Lambda_W) ]
\mathclose{\vert}^2
\label{Z_hgamma}
\end{eqnarray}
where
\begin{equation}
\tau_i = \frac{4 m_i^2}{m_{h^0}^2}, \qquad
\Lambda_i = \frac{4 m_i^2}{m_Z^2}, \qquad
\tau_H = \frac{4 m_H^2}{m_{h^0}^2}, \qquad
\Lambda_H = \frac{4 m_H^2}{m_Z^2},
\label{Lambda_H}
\end{equation}
\begin{eqnarray} 
F(\tau_i, \Lambda_i) & = &
- \frac{1}{2} \frac{\tau_i \Lambda_i}{\tau_i - \Lambda_i}
- \frac{\tau_i^2 \Lambda_i}{\left( \tau_i - \Lambda_i \right)^2}
\left\{ g(\tau_i) - g(\Lambda_i) \right\}
\nonumber \\ & &
+ \frac{1}{4} \frac{\tau_i \Lambda_i}{\tau_i - \Lambda_i}
\left[ 1 + \frac{\tau_i \Lambda_i}{\tau_i - \Lambda_i} \right]
\left\{ f(\tau_i) - f(\Lambda_i) \right\},
\label{F_tau_Lambda}
\end{eqnarray}
\begin{equation}
g(x) = \left\{ \begin{array}{ll}
\left( x - 1 \right)^{1/2} \arcsin{ \left( x^{-1/2} \right) } &
\mbox{if $x \ge 1$} \\
\frac{1}{2} \left( 1 - x \right)^{1/2}
\left[ \ln{ \left\{ \frac{1 + (1 - x)^{1/2}}{1 - (1 - x)^{1/2}} \right\} }
- i \pi \right] &
\mbox{if $x < 1$,}
\end{array} \right.
\label{f1}
\end{equation}
\begin{eqnarray}
I(\tau_H, \Lambda_H) & = &
- \frac{1}{2} \frac{\tau_H \Lambda_H}{(\tau_H - \Lambda_H )}
- \frac{\tau_H^2 \Lambda_H}{(\tau_H - \Lambda_H)^2}
\left[ g(\tau_H) - g(\Lambda_H) \right]
\nonumber \\ & &
+ \frac{1}{4} \frac{\tau_H^2 \Lambda_H^2}{(\tau_H - \Lambda_H)^2}
\left[ f(\tau_H) - f(\Lambda_H) \right],
\label{I2}
\end{eqnarray}
\begin{equation}
\tau_W = \frac{4 m_W^2}{m_{h^0}^2}, \qquad
\Lambda_W = \frac{4 m_W^2}{m_Z^2},
\label{tau_W}
\end{equation}
\begin{equation}
K( \tau_W, \Lambda_W ) = - \frac{\tau_W \Lambda_W}{4 ( \tau_W - \Lambda_W )}
[ f(\tau_W) - f( \Lambda_W ) ].
\label{K}
\end{equation}
The decay width of Equation \ref{Z_hgamma} turns out to be negligible compared to 
the full width of $Z$ so we can not use it to constrain the mass of $h^0$.

\section{Vertex with four particles}
\begin{figure}
\begin{center}
%\vspace*{-4.5cm}
%\scalebox{0.5}
{\includegraphics{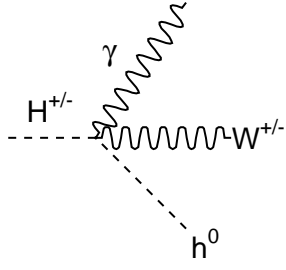}}
%\vspace*{0.7cm}
\caption{Diagram for $H^\pm \rightarrow W^\pm \gamma h^0$.}
\label{hwgammah}
\end{center}
\end{figure}

The decay rate corresponding to the Feynman diagram \ref{hwgammah} is:
\begin{eqnarray}
\lefteqn{
\Gamma \left( H^\pm \rightarrow W^\pm \gamma h^0 \right)
 = \frac{3 G_F^2 \sin^2 \theta_W \left( 1 + \tan \beta \tan \alpha \right)^2
m_W^5}{32 \left( 1 + \tan^2 \beta \right)
\left( 1 + \tan^2 \alpha \right) \pi^3 \left( x_H^W \right)^{1/2}} 
} 
\nonumber \\
& & \times \{ \frac{1}{2} \Lambda^{\frac{1}{2}} \left( 1, x_H^W, x_H^{h^0} \right)
\left( 1 + x_H^W + x_H^{h^0} \right) 
\nonumber \\
& & + \left( 2 x_H^W x_H^{h^0} - x_H^W - x_H^{h^0} \right)
\times \ln \left| \frac{\Lambda^{\frac{1}{2}} \left( 1, x_H^W, x_H^{h^0} \right)
+ 1 - x_H^W - x_H^{h^0}}
{2 \left( x_H^W x_H^{h^0} \right)^{1/2}} \right| 
\nonumber \\
& & - \left| x_H^{h^0} - x_H^W \right| \times
\nonumber \\
& & \ln \left| \frac{ 
\left| x_H^{h^0} - x_H^W \right|
\Lambda^{\frac{1}{2}} \left( 1, x_H^W, x_H^{h^0} \right) 
- \left( x_H^W + x_H^{h^0} \right) +
\left( x_H^W - x_H^{h^0} \right)^2}
{2 \left( x_H^W x_H^{h^0} \right)^{1/2}} \right| \}
\label{H_Wgh}
\end{eqnarray}
where
\begin{equation}
x_H^W = \frac{m_W^2}{m_{H}^2}, \qquad
x_H^{h^0} = \frac{m_{h^0}^2}{m_{H}^2}.
\label{xWHxhH}
\end{equation}

\begin{figure}
\begin{center}
%\vspace*{-4.5cm}
%\scalebox{0.5}
{\includegraphics{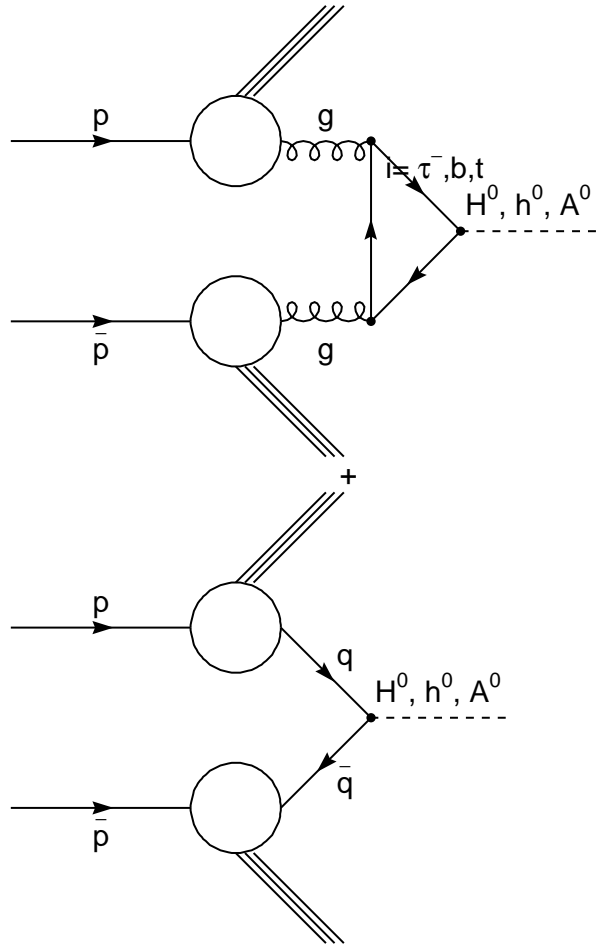}}
%\vspace*{0.7cm}
\caption{Feynman diagrams for 
$p \bar{p} \rightarrow A X$ with
$A \equiv h^0, H^0, A^0$.}
\label{ppbar_A_fig}
\end{center}
\end{figure}

\section{Production of $h^0$, $H^0$ and $A^0$}
From the Feynman diagrams in Figure \ref{ppbar_A_fig} we obtain 
\begin{eqnarray}
\lefteqn{
\sigma \left( p \bar{p} \rightarrow A X \right)  = 
\frac{\pi^2 \Gamma \left( A \rightarrow 2g \right) \gamma_A}
{8 m_A^3}
\int_{\gamma_A}^1 {\frac{dx_a}{x_a} g \left( x_a, m^2 \right) 
g \left( \frac{\gamma_A}{x_a}, m^2 \right)} }
\nonumber \\
& & + \frac{4 \pi^2 \gamma_A}{3 m_A^3}
\left[ \sum_{q=u,d,s,c,b} {\Gamma \left( A \rightarrow q \bar{q} \right)
\int_{\gamma_A}^1 {\frac{dx_a}{x_a} f_q \left( x_a, m^2 \right) 
f_q \left( \frac{\gamma_A}{x_a}, m^2 \right)} } \right]
\label{sigma_pp_AX}
\end{eqnarray}
where $A \equiv h^0, H^0, A^0$ and $\gamma_A \equiv m_A^2/s$.
Here $f_q$ is the unpolarized parton distribution function for quark
or anti-quark
$q$ and $g$ is the parton distribution function for gluons.
$m^2$ is the factorization scale.
$\Gamma(h^0 \rightarrow g g)$ is given by (\ref{2g}),
$\Gamma(H^0 \rightarrow g g)$ by (\ref{H_gg}),
$\Gamma ( A^0 \rightarrow g g )$ by (\ref{A_to_2g}),
$\Gamma\left( h^0 \rightarrow c \bar{c} \right)$ by (\ref{ccbar}),
$\Gamma\left( h^0 \rightarrow b \bar{b} \right)$ by (\ref{bbbar}),
$\Gamma(H^0 \rightarrow q \bar{q})$ by (\ref{H0_ff}),
and finally, $\Gamma \left( A^0 \rightarrow q \bar{q} \right)$ is given
by (\ref{A_ff}).

\section{Production of $h^0 Z^0 X$}
A production channel with interesting experimental signature
is $p \bar{p} \rightarrow h^0 Z^0 X$. The differential cross 
section obtained from the Feynman diagrams in Figure 
\ref{nhiggs2} is
\begin{eqnarray}
\frac{d^2 \sigma}{dy d \left( p_T \right)^2} & = &
\sum_{f} { \int_{x_{amin}}^1 {
dx_a f_f \left( x_a, m_a^2 \right) f_f \left( x_b, m_b^2 \right)
\frac{x_b \hat{s}}{m_{h^0}^2 - \hat{u}} 
}}
\nonumber \\
& & \times \frac{d \sigma}{d \hat{t}} \left( f \bar{f} \rightarrow h^0 Z^0 \right)
\label{pp_hZX}
\end{eqnarray}
where $f$ is $q$ or $g$,
\begin{equation}
x_{amin} = \frac{\sqrt{s} m_T e^y + m_{h^0}^2 - m_Z^2}
{s - \sqrt{s} m_T e^{-y}},
\label{xamin}
\end{equation}
\begin{equation}
m_T = \left( m_Z^2 + p_T^2 \right)^{\frac{1}{2}},
\label{mT}
\end{equation}
\begin{equation}
x_b = \frac{x_a \sqrt{s} m_T e^{-y} + m_{h^0}^2 - m_Z^2}
{x_a s - \sqrt{s} m_T e^y},
\label{xb}
\end{equation}
\begin{equation}
\hat{s} = x_a x_b s,
\label{s_hat}
\end{equation}
\begin{equation}
p_T^2 = \frac{\Lambda \left( \hat{s}, m_{h^0}^2, m_Z^2 \right) \sin^2 \theta}
{4 \hat{s}},
\label{pT2}
\end{equation}
\begin{equation}
\hat{u} = \frac{1}{2} \left( m_{h^0}^2 + m_Z^2 -\hat{s} 
- \cos \theta \Lambda^{1/2} ( \hat{s}, m_{h^0}^2, m_Z^2 ) 
\right)
\label{uhat}
\end{equation}
and
\begin{equation}
\hat{u} \hat{t} = m_{h^0}^2 m_Z^2 + \hat{s} p_T^2.
\label{ut}
\end{equation}
$y$ is the rapidity, $\theta$ is the angle of
dispersion, and $p_T$ is the transverse momentum
of $Z^0$.
For the light quarks $u$, $d$ and $s$ we obtain
\begin{eqnarray}
\frac{d \sigma}{d \hat{t}} & = & \frac{1}{48 \pi \hat{s}}
G_F^2 m_Z^4 \frac{\sin^2 \left( \beta - \alpha \right)}
{\left( \hat{s} - m_Z^2 \right)^2 + m_Z^2 \Gamma_Z^2}
\left[ \left( g_V^f \right)^2 + \left( g_A^f \right)^2 \right]
\nonumber \\
& & \times
\left[ 8 m_Z^2 + \frac{\Lambda \left( \hat{s}, m_{h^0}^2, m_Z^2 \right)}
{\hat{s}} \sin^2 \theta \right]
\label{ds_dthat}
\end{eqnarray}
where $g_A^f \equiv t_{3L} \left( f \right)$ and
$g_V^f \equiv t_{3L} \left( f \right) - 2 q_f \sin^2 \theta_W$.
Coefficients in Equation (\ref{ds_dthat}) are given in Table \ref{c}.
The Standard Model cross section is obtained by
omitting the factor $\sin^2 \left( \beta - \alpha \right)$ in
Equation (\ref{ds_dthat}). The contributions to the cross section
from the heavy quarks $c$ and $b$ are negligible. 
$\Gamma_Z$ is the total decay width of the $Z^0$.
\begin{table}
\begin{center}
\begin{tabular}{|c|c|c|c|c|}
\hline
$f$ & $t_{3L} \left( f \right)$ & $q_f$ & $g_A^f$ & $g_V^f$ \\
\hline
$e^-$, $\mu^-$, $\tau^-$ & $- \frac{1}{2}$ & $-1$ & 
$- \frac{1}{2}$ & $- \frac{1}{2} + 2 \sin^2 \theta_W$ \\
\hline
$u$, $c$, $t$ & $\frac{1}{2}$ & $\frac{2}{3}$ & $\frac{1}{2}$ & 
$\frac{1}{2} - \frac{4}{3} \sin^2 \theta_W$ \\
\hline
$d$, $s$, $b$ & $-\frac{1}{2}$ & $-\frac{1}{3}$ & $-\frac{1}{2}$ &
$-\frac{1}{2} + \frac{2}{3} \sin^2 \theta_W$ \\
\hline
\end{tabular}
\end{center}
\caption{Coefficients in Equation (\ref{ds_dthat}).}
\label{c}
\end{table}

\begin{figure}
\begin{center}
%\vspace*{-4.5cm}
%\scalebox{0.5}
{\includegraphics{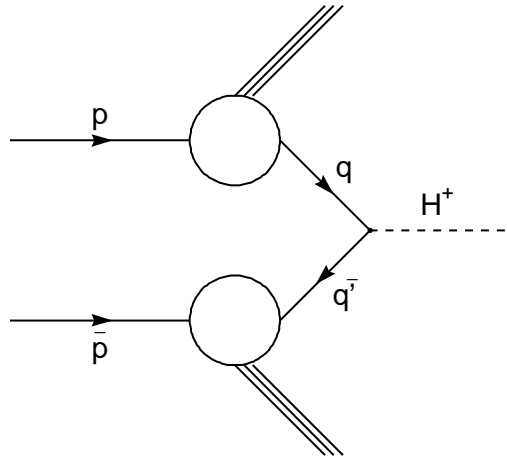}}
%\vspace*{0.7cm}
\caption{Feynman diagram for 
$p \bar{p} \rightarrow H^+ X$.}
\label{ppbar_Hp_fig}
\end{center}
\end{figure}

\begin{figure}
\begin{center}
%\vspace*{-4.5cm}
%\scalebox{0.5}
{\includegraphics{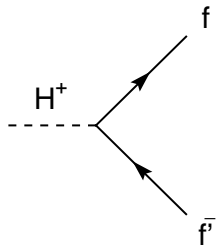}}
%\vspace*{0.7cm}
\caption{Feynman diagram for 
$H^+ \rightarrow f \bar{f}'$.}
\label{Hffbar_fig}
\end{center}
\end{figure}

\section{Production of $H^+$}
From the diagrams in Figures \ref{ppbar_Hp_fig} and \ref{Hffbar_fig} we obtain
\begin{eqnarray}
\lefteqn{
\sigma ( p \bar{p} \rightarrow H^+ X ) = \frac{4 \pi^2 \gamma_H}{3 m_H^3}
\sum_{q, q'}{ \Gamma ( H^+ \rightarrow q \bar{q}' ) }
}
\nonumber \\
& & 
\int_{\gamma_A}^1 {
\frac{dx_a}{x_a} \left[
f_q^p(x_a, m^2) \cdot f_{\bar{q}'}^{\bar{p}} ( \frac{\gamma_H}{x_a}, m^2)
+ f_{\bar{q}'}^p (x_a, m^2) \cdot f_q^{\bar{p}} ( \frac{\gamma_H}{x_a}, m^2 )
\right] }
\label{ppbar_HpX}
\end{eqnarray}
where
\begin{equation}
\gamma_H = \frac{m_H^2}{s}
\label{gamma_H}
\end{equation}
and
\begin{eqnarray}
\lefteqn{
\Gamma (H^+ \rightarrow f \bar{f}' ) = 
\Gamma (H^- \rightarrow f' \bar{f}) }
\nonumber \\
& & = \frac{ \sqrt{2} G_F \left| V_{f f'} \right|^2 N_c}
{16 \pi m_H} \cdot
\Lambda^{1/2} \left( \frac{m_f^2}{m_H^2}, \frac{m_{f'}^2}{m_H^2}, 1 \right)
\nonumber \\
& & \times \left\{ A^2 \left[ m_H^2 - \left( m_f + m_{f'} \right)^2 \right]
+ B^2 \left[ m_H^2 - \left( m_f - m_{f'} \right)^2 \right] \right\}
\label{gamma_H_ff}
\end{eqnarray}
with
$N_c = 3$ for quarks, $N_c = 1$ for leptons,
\begin{equation}
A = m_{f'} \tan \beta + m_f \cot \beta
\label{AAA}
\end{equation}
\begin{equation}
B = m_{f'} \tan \beta - m_f \cot \beta
\label{B}
\end{equation}
$f = u, c, t, \nu_e, \nu_\mu, \nu_\tau$,
and $f' = d, s, b, e^-, \mu^-, \tau^-$.\cite{hernandez}

\section{Production of $h^0 W^+ X$}
Let us now consider the channel $p \bar{p} \rightarrow h^0 W^+ X$.
We obtain:
\begin{eqnarray}
\frac{d^2 \sigma}{dy d \left( p_T \right)^2} & = &
\sum_{q, q'} { \int_{x_{amin}}^1 {
dx_a [ f_{q}^p \left( x_a, m_a^2 \right) f_{\bar{q'}}^{\bar{p}} 
\left( x_b, m_b^2 \right) }}
\nonumber \\
& & + f_{\bar{q'}}^p \left( x_a, m_a^2 \right) f_{q}^{\bar{p}}
\left( x_b, m_b^2 \right) ]
\frac{x_b \hat{s}}{m_{h^0}^2 - \hat{u}} 
%\nonumber \\
%& & \times 
\frac{d \sigma}{d \hat{t}} \left( q  \bar{q'} \rightarrow h^0 W^+ \right)
%}}
\label{pp_hWX}
\end{eqnarray}
where
\begin{equation}
f_{q'}^p = f_{\bar{q'}}^p, \qquad
f_{\bar{q}}^{\bar{p}} = f_q^{\bar{p}}, \qquad
f_{\bar{q}}^p = f_q^p, \qquad
f_{q'}^{\bar{p}} = f_{\bar{q'}}^{\bar{p}},
\label{ff}
\end{equation}
\begin{equation}
x_{amin} = \frac{\sqrt{s} m_T e^y + m_{h^0}^2 - m_W^2}
{s - \sqrt{s} m_T e^{-y}},
\label{xamin2}
\end{equation}
\begin{equation}
m_T = \left( m_W^2 + p_T^2 \right)^{\frac{1}{2}},
\label{mT2}
\end{equation}
\begin{equation}
x_b = \frac{x_a \sqrt{s} m_T e^{-y} + m_{h^0}^2 - m_W^2}
{x_a s - \sqrt{s} m_T e^y},
\label{xb2}
\end{equation}
\begin{equation}
\hat{s} = x_a x_b s,
\label{s_hat2}
\end{equation}
\begin{equation}
p_T^2 = \frac{\Lambda \left( \hat{s}, m_{h^0}^2, m_W^2 \right) \sin^2 \theta}
{4 \hat{s}},
\label{pT22}
\end{equation}
\begin{equation}
\hat{u} = \frac{1}{2} \left[ m_{h^0}^2 + m_W^2 -\hat{s} 
- \cos \theta \Lambda^{1/2} ( \hat{s}, m_{h^0}^2, m_W^2 ) 
\right],
\label{uhat2}
\end{equation}
\begin{equation}
\hat{t} = \frac{1}{2} \left[ m_{h^0}^2 + m_W^2 -\hat{s} 
+ \cos \theta \Lambda^{1/2} ( \hat{s}, m_{h^0}^2, m_W^2 ) 
\right],
\label{that2}
\end{equation}
\begin{equation}
\cos \theta = \left( 1 -
\frac{4 \hat{s} p_T^2}{\Lambda (\hat{s}, m_{h^0}^2, m_W^2 )} \right)^{1/2}
\label{costheta2}
\end{equation}
and
\begin{equation}
\hat{u} \hat{t} = m_{h^0}^2 m_W^2 + \hat{s} p_T^2.
\label{ut2}
\end{equation}
$y$ is the rapidity of $W^+$ and $p_T$ is the transverse momentum of $W^+$.
From the Feynman diagrams of Figure \ref{wh_fig} we obtain
for $f \bar{f'} \rightarrow h^0 W^+$:
\begin{eqnarray}
\lefteqn{
\frac{d \sigma}{d \hat{t}}  =  \frac{1}{16 \pi \hat{s}^2}
\left| V_{f f'} \right|^2 G_F^2
\{ 
\left| C_{H^+} \right|^2 \hat{s}
\Lambda 
}
\nonumber \\ & & 
\times \left[ m_{f'}^2 \tan^2 \beta + m_f^2 \cot^2 \beta \right]
+ m_W^4 \left| C_W \right|^2 
\left[ 8 \hat{s} m_W^2 + \Lambda \sin^2 \theta \right]
\nonumber \\ & & 
- 2 C_{H^+} \Re{ (C_W) }
[ m_{f'}^2 \tan \beta ( \hat{s} \Lambda 
+ 2 m_W^2 \hat{u} \left( \hat{s} - m_{h^0}^2 \right)
\nonumber \\ & &
+ 2 m_W^4 \left( 2 m_{h^0}^2 - \hat{t} \right) )
- m_f^2 \cot \beta ( \hat{s} \Lambda + 2 m_W^2 \hat{t}
\left( \hat{s} - m_{h^0}^2 \right) 
\nonumber \\ & &
+ 2 m_W^4 \left( 2 m_{H^0}^2 - \hat{u} \right) 
) ]
\nonumber \\ & &
+ \frac{1}{2} m_W^2 \Lambda \sin^2 \theta
\left[ \frac{m_f^2 C_f^2}{\hat{t}^2} + \frac{m_{f'}^2 C_{f'}^2}{\hat{u}}
\right] +
\hat{s} \left[ m_f^2 C_f^2 + m_{f'}^2 C_{f'}^2 \right]
\nonumber \\ & &
+ 2 C_{H^+} \hat{s} 
\left[ m_{h^0}^2 m_W^2 - \frac{1}{4} \Lambda \sin^2 \theta \right]
\left[ \frac{m_f^2 \cot \beta C_f}{\hat{t}} - 
\frac{m_{f'}^2 \tan \beta C_{f'}}{\hat{u}} \right]
\nonumber \\ & &
+ 2 C_{H^+} \hat{s}
\left[ m_{f'}^2 \tan \beta C_{f'} \hat{u} - m_f^2 \cot \beta C_f \hat{t} \right]
\nonumber \\ & &
-2 \Re{(C_W)} \left[ \frac{1}{2} \Lambda \sin^2 \theta
\left( m_W^2 + \frac{\hat{s}}{2} \right)
- \hat{s} m_{h^0}^2 m_W^2 + 4 \hat{s} m_W^4 + 4 m_W^6 \right]
\nonumber \\ & &
\times \left[ \frac{m_f^2 C_f}{\hat{t}} + \frac{m_{f'}^2 C_{f'}}{\hat{u}} \right]
- 2 \Re{(C_W)} m_f^2 C_f \left[ -2 m_W^4 + \hat{t} 
\left( \hat{s} - 2 m_W^2 \right) \right]
\nonumber \\ & &
- 2 \Re{(C_W)} m_{f'}^2 C_{f'} \left[ -2 m_W^4 + \hat{u} 
\left( \hat{s} - 2 m_W^2 \right) \right]
\}
\label{pp_h0W}
\end{eqnarray}
where $\Lambda$ stands for $\Lambda ( \hat{s}, m_{h^0}^2, m_W^2 )$,
\begin{equation}
C_{H^+} = \frac{\cos \left( \beta - \alpha \right)}{\hat{s} - m_H^2},
\label{CHp}
\end{equation}
\begin{equation}
C_W = \frac{\sin \left( \beta - \alpha \right)
\left( \hat{s} - m_W^2 - i m_W \Gamma_W \right)}
{\left( \hat{s} - m_W^2 \right)^2 + m_W^2 \Gamma_W^2},
\label{CW}
\end{equation}
\begin{equation}
C_f = - \frac{\cos \alpha}{\sin \beta}, \qquad
C_{f'} = \frac{\sin \alpha}{\cos \beta}.
\label{Cf}
\end{equation}
For $p \bar{p} \rightarrow h^0 W^- X$ interchange
$\hat{u} \leftrightarrow \hat{t}$.

For the Standard Model we obtain the differential
cross section (\ref{pp_h0W}) with $h^0$ replaced
by the Standard Model higgs, $C_{H^+} = 0$, 
$C_f = C_{f'} = -1$, and $\sin(\beta - \alpha) = 1$ in (\ref{CW}).

\section{Numerical examples}
Two sensitive channels for the search of the Standard Model higgs
are $p \bar{p} \rightarrow h^0 Z X$ and
$p \bar{p} \rightarrow h^0 W^\pm X$.
The cross section for $p \bar{p} \rightarrow h^0 Z X$
off resonance
in the Doublet model differs from the Standard Model
by a factor $\sin^2 (\beta - \alpha)$
(see Equation (\ref{ds_dthat}))
and it will
be hard to obtain both $m_{h^0}$ and
$\tan(\beta)$. We are therefore interested
in the production of $h^0 Z$ on resonance.
In particular
$p \bar{p} \rightarrow A^0$ followed by
$A^0 \rightarrow h^0 Z \rightarrow b \bar{b} l^- l^+$
where $l = \mu, e$. A peak should be observed in
the $h^0 Z$ invariant mass.
From Equation (\ref{sigma_pp_AX})
we obtain the cross sections listed in Tables
\ref{numbers_A0_200} and \ref{numbers_A0_250}.

\begin{table}
\begin{center}
\begin{tabular}{|c|c|c|c|}
\hline
partons & $\tan(\beta) = 100$ & $\tan(\beta) = 10$ & $\tan(\beta) = 2$ \\
\hline
$g g$       & 0.20E+1 & 0.13E-1 & 0.35E-1 \\
$b \bar{b}$ & 0.31E+2 & 0.31E+0 & 0.12E-1 \\
$c \bar{c}$ & 0.73E-7 & 0.73E-5 & 0.18E-3 \\
$s \bar{s}$ & 0.20E+0 & 0.20E-2 & 0.81E-4 \\
$d \bar{d}$ & 0.10E-1 & 0.10E-3 & 0.41E-5 \\
$u \bar{u}$ & 0.18E-9 & 0.18E-7 & 0.46E-6 \\
\hline
\end{tabular}
\end{center}
\caption{Production cross section [pb] for
$p \bar{p} \rightarrow A^0$ from the indicated
partons. $m_{A^0} = 200$GeV/c$^2$, $\sqrt{s} = 1960$GeV/c$^2$.}
\label{numbers_A0_200}
\end{table}

\begin{table}
\begin{center}
\begin{tabular}{|c|c|c|c|}
\hline
partons & $\tan(\beta) = 100$ & $\tan(\beta) = 10$ & $\tan(\beta) = 2$ \\
\hline
$g g$       & 0.42E+0 & 0.22E-2 & 0.19E-1 \\
$b \bar{b}$ & 0.82E+1 & 0.82E-1 & 0.33E-2 \\
$c \bar{c}$ & 0.19E-7 & 0.19E-5 & 0.49E-4 \\
$s \bar{s}$ & 0.57E-1 & 0.57E-3 & 0.23E-4 \\
$d \bar{d}$ & 0.44E-2 & 0.44E-4 & 0.18E-5 \\
$u \bar{u}$ & 0.89E-10 & 0.89E-8 & 0.22E-6 \\
\hline
\end{tabular}
\end{center}
\caption{Production cross section [pb] for
$p \bar{p} \rightarrow A^0$ from the indicated
partons. $m_{A^0} = 250$GeV/c$^2$, $\sqrt{s} = 1960$GeV/c$^2$.}
\label{numbers_A0_250}
\end{table}

Let us now consider the decays of $A^0$. As an example
we take $m_{h^0} = 120$GeV/c$^2$, $m_{H^0} = 250$GeV/c$^2$,
$m_H = 200$GeV/c$^2$ and $m_{A^0} = 250$GeV/c$^2$.
The corresponding branching fractions are listed in
Table \ref{BR_A0}. From Tables \ref{numbers_A0_250} and
\ref{BR_A0} we obtain a production cross section times
branching fraction for the process
$p \bar{p} \rightarrow A^0 \rightarrow h^0 Z$
of 0.018pb for $\tan(\beta) = 2$, and
0.0045pb for $\tan(\beta) = 10$.

\begin{table}
\begin{center}
\begin{tabular}{|c|c|c|c|}
\hline
partons & $\tan(\beta) = 100$ & $\tan(\beta) = 10$ & $\tan(\beta) = 2$ \\
\hline
$A \rightarrow gg$            & 3.0E-4 & 1.5E-4 & 2.0E-3 \\
$A \rightarrow b \bar{b}$     & 1.0E+0 & 9.4E-1 & 5.9E-2 \\
$A \rightarrow c \bar{c}$     & 8.0E-10 & 7.5E-6 & 2.9E-4 \\
$A \rightarrow s \bar{s}$     & 8.0E-4 & 7.5E-4 & 4.7E-5 \\
$A \rightarrow Z h^0$         & 6.1E-6 & 5.5E-2 & 9.4E-1 \\
$A \rightarrow Z \gamma$      & 1.9E-8 & 6.0E-9 & 1.2E-6 \\
$A \rightarrow \gamma \gamma$ & 1.2E-7 & 8.1E-8 & 7.5E-6 \\
\hline
\end{tabular}
\end{center}
\caption{Branching fractions for $A^0$ assuming
$m_H = 200$GeV/c$^2$, $m_{H^0} = 250$GeV/c$^2$, $m_{A^0} = 250$GeV/c$^2$
and $m_{h^0} = 120$GeV/c$^2$.}
\label{BR_A0}
\end{table}

From Equations (\ref{ppbar_HpX}) and (\ref{gamma_H_ff}) we obtain the production
cross sections for $p \bar{p} \rightarrow H^+ X$ shown in
Table \ref{sigmaH}.
\begin{table}
\begin{center}
\begin{tabular}{|c|c|c|c|}
\hline
partons & $\tan(\beta) = 100$ & $\tan(\beta) = 10$ & $\tan(\beta) = 2$ \\
\hline
$u \bar{d}$ & 0.82E-2 & 0.82E-4 & 0.34E-5 \\
$u \bar{s}$ & 0.66E-1 & 0.66E-3 & 0.26E-4 \\
$u \bar{b}$ & 0.89E-2 & 0.89E-4 & 0.36E-5 \\
$c \bar{s}$ & 0.31E-1 & 0.32E-3 & 0.91E-4 \\
$c \bar{b}$ & 0.24E-1 & 0.24E-3 & 0.96E-5 \\
\hline
\end{tabular}
\end{center}
\caption{Production cross section [pb] for
$p \bar{p} \rightarrow H^+ X$ from the indicated
partons. $m_{H^0} = 250$GeV/c$^2$, $\sqrt{s} = 1960$GeV/c$^2$.}
\label{sigmaH}
\end{table}

\begin{figure}
\begin{center}
%\vspace*{-4.5cm}
%\scalebox{0.5}
{\includegraphics{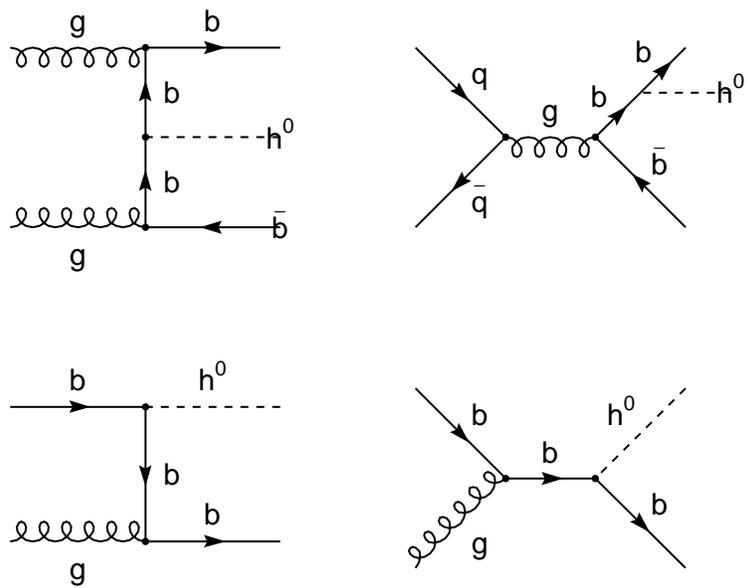}}
%\vspace*{0.7cm}
\caption{Some Feynman diagrams for the production of
three or more $b$-jets.}
\label{gluonb_fig}
\end{center}
\end{figure}

Other channels of experimental interest are the production of
3 or more $b$-jets as in Figure \ref{gluonb_fig}. Some numerical calculations using the 
CompHEP program\cite{comphep} are presented in Table \ref{comphep_tab}.

\begin{table}
\begin{center}
\begin{tabular}{|c|c|c|c|}
\hline
process & $\tan(\beta) = 100$ & $\tan(\beta) = 50$ & $\tan(\beta) = 2$ \\
\hline
$b +  g \rightarrow b + h^0$ & 0.021 & 0.021 & 0.011 \\
$u + \bar{u} \rightarrow b + \bar{b} + h^0$ & 0.002 & 0.001 & 0.0004 \\
$d + \bar{d} \rightarrow b + \bar{b} + h^0$ & 0.0005 & 0.0005 & 0.0001 \\
$g + g \rightarrow b + \bar{b} + h^0$ & 0.015 & 0.015 & 0.008 \\
\hline
\end{tabular}
\end{center}
\caption{Production cross section [pb] for
$p \bar{p} \rightarrow b h^0 X$ from the indicated
processes. $m_{h^0} = 120$GeV/c$^2$, 
$m_{A^0} = 250$GeV/c$^2$,
$\sqrt{s} = 1960$GeV/c$^2$.}
\label{comphep_tab}
\end{table}

\section{Running coupling constants and Grand Unification}
The coupling constants of the Two Higgs Doublet Model of type II
are $g_s(\mu)$ for SU(3), $g(\mu)$ for SU(2), and $g'(\mu)$ for
U(1). 
These coupling constants
depend on the energy scale $\mu$ as follows:
\begin{equation}
\frac{1}{g_s^2(\mu)} = \frac{1}{g_s^2(m_x)} + 
\frac{1}{8 \pi^2} \left( -11 + \frac{4}{3} n_F \right) 
\ln{ \left( \frac{m_x}{\mu} \right) },
\label{gs}
\end{equation}
\begin{equation}
\frac{1}{g^2(\mu)} = \frac{1}{g^2(m_x)} + 
\frac{1}{8 \pi^2} \left( - \frac{22}{3} + \frac{4}{3} n_F + \frac{1}{6} n_S\right) 
\ln{ \left( \frac{m_x}{\mu} \right) },
\label{g}
\end{equation}
\begin{equation}
\frac{1}{g'^2(\mu)} = \frac{1}{g'^2(m_x)} + 
\frac{1}{8 \pi^2} \left( \frac{20}{9} n_F + \frac{1}{6} n_S \right) 
\ln{ \left( \frac{m_x}{\mu} \right) },
\label{g_prime}
\end{equation}
where $n_F$ is the number of 
families of quarks and leptons, and $n_S$ is the number of
higgs doublets. For the Two Higgs Doublet Model of type II considered
in this article, $n_F = 3$ and $n_S = 2$. In terms of the
elementary electric charge and the Weinberg angle,
$g(m_Z) = e(m_Z)/\sin \theta_W(m_Z)$, 
$g'(m_Z) = e(m_Z)/\cos \theta_W(m_Z)$. The fine structure 
constant is $\alpha(m_Z) = e^2(m_Z)/(4 \pi)$.

Let us now assume that a Grand Unified Theory (GUT) breaks its symmetry
to SU(3)$\times$SU(2)$\times$U(1) at the energy scale $m_x$.
At this scale we take
\begin{equation}
g_s^2(m_x) = g^2(m_x) = \frac{5}{3} g'^2(m_x),
\label{gmx}
\end{equation}
and obtain
\begin{equation}
\sin^2 \theta_W = \frac{11 + \frac{1}{2} n_S + \frac{5e^2}{3g_s^2} 
\left( 22 - \frac{1}{5} n_S \right)}
{ 66 + n_S },
\label{sin2tW}
\end{equation}
\begin{equation}
\ln{ \left( \frac{m_x}{m_Z} \right) } =
\frac{24 \pi^2}{e^2}
\frac{1 - \frac{8e^2}{3g_s^2} }{66 + n_S},
\label{mx}
\end{equation}
with all running couplings evaluated at $m_Z$.

The corresponding equations of the Minimum Supersymmetry Model\cite{MSSM}
are
\begin{equation}
\sin^2 \theta_W = \frac{18 + 3n_S + \frac{e^2}{g_s^2} 
\left( 60 - 2n_S \right) }
{108 + 6n_S},
\label{sin2tW_MSSM}
\end{equation}
\begin{equation}
\ln{ \left( \frac{m_x}{m_Z} \right) } =
\frac{8 \pi^2}{e^2}
\left[ \frac{1 - \frac{8e^2}{3g_s^2} }{18+n_S} \right].
\label{mx_MSSM}
\end{equation}

Some numerical results are presented in Table \ref{sint_mx}.
From the Table we conclude that the Two Higgs Doublet Model 
of type II is in disagreement with the measured value
of $\sin^2 \theta_W(m_Z)$, and with the non-observation of
proton decay ($m_x$ is too low). Raising the number of doublets to $\approx 7$
would bring $\sin^2 \theta_W(m_Z)$ into agreement with 
observations, but $m_x$ is still too low. The
MSSM with $n_S = 2$ (which includes the Two Higgs Doublet Model
of type II) is in agreement with both the observed
$\sin^2 \theta_W(m_Z)$, and with the non-observation of
proton decay.

\begin{table}
\begin{center}
\begin{tabular}{|c|c|c|c|c|}
\hline
 & \multicolumn{2}{|c|}{Doublet Model}
 & \multicolumn{2}{|c|}{MSSM} \\
%%\cline{2-5}
\hline
$n_S$ & $\sin^2 \theta_W(m_Z)$ & $m_x$ & $\sin^2 \theta_W(m_Z)$ & $m_x$ \\
\hline
0 & 0.2037 & $1.0\cdot10^{15}$ & 0.2037 & $8.0\cdot10^{17}$ \\ 
2 & 0.2118 & $4.2\cdot10^{14}$ & 0.2311 & $2.0\cdot10^{16}$ \\ 
4 & 0.2194 & $1.8\cdot10^{14}$ & 0.2536 & $1.0\cdot10^{15}$ \\ 
6 & 0.2266 & $8.3\cdot10^{13}$ & 0.2722 & $8.3\cdot10^{13}$ \\ 
8 & 0.2334 & $3.9\cdot10^{13}$ & 0.2880 & $1.0\cdot10^{13}$ \\ 
\hline
\end{tabular}
\end{center}
\caption{Predicted $\sin^2 \theta_W(m_Z)$ and $m_x$ for the
Two Higgs Doublet Model of type II, and the Minimum Supersymmetric
Model as a function of the number of doublets $n_S$.}
\label{sint_mx}
\end{table}

\section{Conclusions}
One of the major efforts at the Fermilab
Tevatron in Run II, and at the future LHC, is the search for the
Standard Model higgs $h_{SM}$. The four channels with largest
production cross section are\cite{PDG}
$gg \rightarrow h_{SM}$, $q \bar{q'} \rightarrow h_{SM} W$,
$q \bar{q} \rightarrow h_{SM} Z$ and $qq \rightarrow h_{SM} qq$.
The decay modes of $h_{SM}$ with largest branching fraction\cite{PDG}
are $b \bar{b}$ for $m_h \lesssim 137$GeV and $W^+ W^-$
for $m_h \gtrsim 137$GeV.

The search for the Standard Model higgs will also constrain 
or discover particles of the
Two Higgs Doublet Model of type II. 

The most interesting production channels are
$gg \rightarrow h^0, H^0, A^0$ on mass shell, and
$q \bar{q}, g g \rightarrow h^0 Z$ and
$q \bar{q'} \rightarrow h^0 W^\pm$ 
in the continuum
(tho there may be peaks at $m_{A^0}$).
The most interesting decays are
$h^0, H^0, A^0 \rightarrow b \bar{b}$-jets and
$\tau^+ \tau^-$, and, if above threshold,
$H^0 \rightarrow 
Z Z$, $W^+ W^-$ and $h^0 h^0$.
The following final states should be compared with the 
Standard Model cross section:
$b \bar{b} Z$, $b \bar{b} W^\pm$, 
$\tau^+ \tau^- Z$, $\tau^+ \tau^- W^\pm$,
$b \bar{b}$, $\tau^+ \tau^-$, $Z Z$,
$W^+ W^-$, 3 and 4 $b$-jets, $2 \tau^+ + 2 \tau^-$,
$b \bar{b} \tau^+ \tau^-$, $Z W^+ W^-$, $3 Z$,
$Z Z W^\pm$ and $3 W^\pm$.
Mass peaks should be searched in the following
channels:
$Z b \bar{b}$, $Z Z$, $Z Z Z$, $b \bar{b}$, $4 b$-jets 
and, just in case, $Z \gamma$.

We have discussed the masses of the higgs
particles in the Two Higgs Doublet Model of type II, and have
calculated several relevant production and decay rates.
We have also discussed running coupling constants and
Grand Unification. If the Two Higgs Doublet Model of type II
is part of a Grand Unified Theory, then it does not
agree with the observed $\sin^2 \theta_W$ nor with the
non-observation of proton decay. The
MSSM with $n_S = 2$ (which includes the Two Higgs Doublet Model
of type II) is in agreement with both the observed
$\sin^2 \theta_W(m_Z)$, and with the non-observation of
proton decay.

\end{document}